\documentclass{article}
\usepackage[utf8]{inputenc}
\usepackage[margin=3.5cm]{geometry}
\usepackage[numbers]{natbib}
\usepackage{graphicx}
\usepackage{todonotes}
\usepackage{amsmath,bm,amsfonts,amssymb}
\usepackage{amsthm}
\usepackage{float}
\usepackage{graphicx}
\usepackage{multirow,array}
\usepackage{subfig}
\usepackage{colortbl}
\usepackage{authblk}
\usepackage{soul}
\newtheorem*{remark}{Remark}

\graphicspath{{figures/}}

\title{On Calibration Neural Networks for extracting implied information from American options}

\author[1]{\small Shuaiqiang Liu}
\author[2,3]{\small \'Alvaro Leitao}
\author[4]{\small Anastasia Borovykh}
\author[1,5]{\small Cornelis W. Oosterlee}

\affil[1]{\footnotesize Applied Mathematics (DIAM), Delft University of Technology, Delft, the Netherlands}
\affil[2]{\footnotesize Department of Mathematics, University of A Coruña, A Coruña, Spain}
\affil[3]{\footnotesize CITIC research centre, University of A Coruña, A Coruña, Spain}
\affil[4]{\footnotesize  Imperial College London, UK}
\affil[5]{\footnotesize Centrum Wiskunde \& Informatica, Amsterdam, the Netherlands}
\date{}

\begin{document}

\maketitle
\begin{abstract}
Extracting implied information, like volatility and/or dividend, from observed option prices is a challenging task when dealing with American options, because of the computational costs needed to solve the corresponding mathematical problem many thousands of times.
We will employ a data-driven machine learning approach to estimate the Black-Scholes implied volatility and the dividend yield for American options in a fast and robust way. To determine the implied volatility, the inverse function is approximated by an artificial neural network on the computational domain of interest,
which decouples the offline (training) and online (prediction) phases and thus eliminates the need for an iterative process.
For the implied dividend yield, we formulate the inverse problem as a calibration problem and determine simultaneously the implied volatility and dividend yield.
For this, a generic and robust calibration framework, the Calibration Neural Network (CaNN), is introduced to estimate multiple parameters.
It is shown that machine learning can be used as an efficient numerical technique to extract implied information from American options.

\end{abstract}

\section{Introduction}
American options give the holder the right (but not the obligation) to exercise the contract at any time before the contract's expiry. American-style options are very common in the market, and the underlying asset may be any financial asset, such as currency, commodity, or stock.
Computing American-style option prices is generally more expensive than pricing European-style options, because there is no closed-form expression.

The inverse problem, i.e. determining implied information from observed American option prices, which traditionally requires solving the pricing model many thousands of times, is a highly challenging task. Recently, deep neural networks have emerged as an advanced computational technique to efficiently obtain solutions to possibly high-dimensional problems in computational finance, e.g.  option pricing~\cite{Han8505,risks7010016,lokeshwar2019neural}, model calibration~\cite{CaNN2019}, hedging~\cite{bhler2018deep}, optimal stopping~\cite{optimalstopping2019}. We refer to~\cite{ruf2019neural} for a review.

Implied volatility represents a specific measure of the future uncertainty in stock prices, from the market point of view.  Computing the implied volatility from American options requires inverting the American-style pricing model, where the main difficulty is solving the well-known free boundary problem.
Other complicating factors, such as a negative interest rate~\cite{Frankena2016PricingAH} or dividend yield~\cite{DoubleContinuation2015}, may lead to complex-shaped early-exercise regions (e.g.  we may encounter so-called two continuation regions)\cite{Doublecontinuation2018}.

There are essentially two popular ways of determining an American option implied volatility. The first one is by means of a de-Americanization method, which translates an American option price into the corresponding European version~\citep{deAmReview2018,optionswaps2009,CalbrationAm2019}.
However, significant pricing errors  may  arise due to inaccurate incorporation of the early-exercise premium.
 The higher the early-exercise premium, typically  the larger the error can get. Taking dividends into consideration may further increase the error of the de-Americanization method. The second approach is to conduct a direct calibration for the American-style pricing model to extract implied volatility, see~\citep{achdou2004, CalbrationAm2019,1998AmVolQam,OsherCalibrating1997}.
More precisely, computing the implied volatility from American options is defined as a minimization problem, using an iterative search technique to repeatedly compare model with market option prices, until the suitable value of volatility is found. The above two methods thus rely on  an iterative numerical procedure. However, valuation of American options requires intensive computation and the boundaries of the early-exercise regions are not known in advance.
Unlike the European-style options,  the derivative of the option value with respect to the volatility does not have a closed-form expression in the case of American-style options.
Besides intensive computation, the search technique may fail to converge when it accidentally explores the early-exercise region (where the output appears insensitive to input parameter changes, as the gradient, Vega, equals zero in that region).

The implied dividend is a measure which reflects how much an underlying asset price will be reduced over the option's lifetime under the risk-neutral measure. Compared to implied volatility, the implied dividend is typically a relatively weak signal, but there are several applications for the implied dividend. First of all,
an accurate implied dividend yield may result in accurate theoretical option values, Greeks and implied volatilities. Moreover,  a hedging process involves frequent updating of the position in the underlying asset, and one may choose a specific trading strategy according to the difference between market implied and  announced actual dividends, see~\cite{J.Hull}.
The authors in~\cite{CutsFromimpliedDiv2017, ForecaswithOptionImpliedinfo2013} give evidence for the fact that implied dividends are a significant factor for forecasting actual dividend changes.
Implied dividend is shown to have a more predictive power than historical dividends.  In~\cite{orat}, a dividend forecast model is developed based on market implied dividend, together with historical dividend, dividend seasonal effects and some fundamental analysis.

A popular way to estimate implied dividend is by means of the put-call parity. The put-call parity holds, however, only for European-style options, and does no longer hold in the case of American-style options  due to the early-exercise premium~\cite{J.Hull,CutsFromimpliedDiv2017}. Alternatively, an option pricing model including a dividend yield model can be inverted to determine the implied dividend, as in~\cite{amdiv2013} where a known implied volatility was assumed. Usually, however, implied volatility is also unknown, so that there are two unknowns to determine in calibration, see for
example,~\cite{Qi2005, OptionImpliedDividends2017}.

For computing both implied volatility and implied dividend, these unknowns will be determined based on two pricing functions.  With a suitable pair of market American call/put prices, the system should produce a unique solution mathematically. Moreover, multiple option price quotes may share the same implied dividend~\cite{Qi2005}, and an over-determined system may arise in such case. In order to handle the various situations (e.g. more than two prices) in a unified framework, a robust and efficient calibration framework needs to be employed to determine the implied volatility and implied dividend simultaneously, which is named Calibration Neural Network, CaNN~\cite{CaNN2019}. More specifically, CaNN is composed of three components, an efficient option pricing method, a global optimization technique and an implementation which runs efficiently on a parallel computing platform (e.g. GPU). There are two separate stages in CaNN, the forward pass to learn the pricing model and the backward pass to estimate the model parameters. Here the forward pass of CaNN will give us two output quantities,  originating from one neural network. Possible negative interest rates and a negative dividend yield are taken into consideration to cover a broad variety of market conditions.

When the dividend is known, we will compute American-style implied volatility by approximating the inverse function with an artificial neural network (ANN), i.e. not requiring an iterative numerical method. For American-style options, it does not make sense to invert the pricing model in the early-exercise region. This means that the definition domain of the inverse function is the option's continuation region (whose boundaries are not known in advance). ANNs can approximate a function in a complex geometry, see, for example, \citep{Berg2018AUD}. We will numerically determine the free boundaries during the off-line learning phase. Although this procedure needs additional computation, it won't affect the solution time during the prediction phase.  Thus,  computing the implied volatility can be done swiftly during the on-line phase.

The remainder of this paper is organized as follows. In Section \ref{section:American options}, the mathematical American option pricing models are introduced. Furthermore, in Section \ref{section:American implied info}, the implied volatility and implied dividend yield from American options are discussed.
In Section \ref{section:Methodology}, we  describe the data-driven ANN to extract implied information from American options. When the dividend yield is known, we can use the ANN to approximate the inverse function. In other cases, the CaNN is employed to determine both the implied dividend and implied volatility.  In Section \ref{section:numerical results}, numerical experiments are presented to demonstrate the performance of the proposed methods.

%---------------------------------

%-----
\section{American options} \label{section:American options}
In this section we will discuss the mathematical model used to price the American-style options, as well as the implied information from market option prices.

\subsection{Problem formulation}
Although other pricing models would easily fit into our framework, for clarity we will concentrate on the Black-Scholes pricing framework. The underlying asset price thus follows a Geometric Brownian Motion (GBM) process, under the risk-neutral measure,
\begin{equation} \label{eq:stockmodel}
    dS_t = S_t (r-q) dt +  \sigma S_t dW_t,
\end{equation}
where $S_t$ is the underlying spot price, and $\sigma$ is the volatility parameter and $W_t$ a Wiener process. The two parameters $r$ and $q$ can be interpreted in different ways. For example, $r$ and $q$ are the risk-less interest rate and dividend yield, respectively, for stocks. In the context of currencies, $r$ and $q$ may be two different interest rates, and $q$ would represent the cost of carry in the case of commodities. In this paper, we stay with stock option descriptions, for convenience, but we will also discuss $q<0$, which is found in commodity modeling.  With a risk-less asset $B_t$,
\begin{equation} \label{eq:cashmodel}
    dB_t = r B_t dt,
\end{equation}
the arbitrage-free value of an American option  at time $t$ is given by
\begin{equation} \label{eq:optionmodel}
    V_{am}(S_t, t)  = \sup \limits_{u \in [0,T]} \mathbb{E}_t^\mathbb{Q}[e^{-r(T-t)}H(K, S_u) | S_u],
\end{equation}
where $H(\cdot)$ is the payoff function, with strike price $K$; $\mathbb{E}_t^\mathbb{Q}$ represents the expectation under the risk-neutral measure $\mathbb{Q}$, expiration time is $T$.
An optimal exercise boundary $S^*_t \equiv S^*(t)$, which depends on the time to maturity $T-t$, divides the domain into early-exercise (stopping) regions $\Omega_s$ and continuation (or holding) regions $\Omega_h$.
In general, early-exercise will be triggered when the discounted expected value drops below the value of exercising the option.
As an American option can be exercised anytime before the expiration time, a corresponding early-exercise premium should be added to the European option counterpart.
An American put option~\citep{earlypremium1992} can be decomposed into the corresponding European put price and the early-exercise premium,
\begin{equation}\label{eq:AmPutIntegral}
    V^P_{am}(S_t, t) = \mathbb{E}^\mathbb{Q}_t[e^{-r(T-t)} \max(K-S_T,0)] + \int_t^T \mathbb{E}^\mathbb{Q}_u[(rK-qS_u)\textbf{1}_{\{S_u \in \Omega_s\}}]du,
\end{equation}
where $\Omega_s$ represents the stopping region, and the whole domain is $\Omega=\Omega_s+\Omega_h$ with $\Omega_h$ being the holding region.

The first term in Equation \eqref{eq:AmPutIntegral} is indeed equivalent to the European Black-Scholes put solution.
The problem can be formulated  as a Black-Scholes inequality, on a domain $\Omega$ with a free boundary $S^*_t$.
At the free boundary, we have,
\begin{equation} \label{eq:freeboundary}
    V_{am}(S^*,t) = H(K,S_t^*), \;\;\; \frac{\partial V_{am}}{ \partial S} = \alpha,
\end{equation}
where $\alpha=1$ for American calls, $\alpha=-1$ for American puts.  $S_t^*$ is the asset price for which the option value equals the payoff function. For American put options,  the payoff function equals $H(K,S_t) = \max(K-S_t,0)$.  The above conditions at the free boundary can be used to distinguish continuation from stopping regions, and we will use these conditions in Section \ref{section: iv-ann-method}. In this paper, the American Black-Scholes solution and corresponding pricing model are denoted by $V_{am}=BS_{am}(\sigma,S_0,K,\tau,r,q,\alpha)$, with $\tau:=T-t$.

%----------------
\subsection{Put-call symmetry}
The {\em put-call symmetry} relation allows us to value a call or put option by means of its counterpart, where the role of stock and cash are exchanged. The put-call symmetry relation holds for both European options and American options, and is given by,
 \begin{equation} \label{eq:pcsy}
     V^P(K,t,T,S,\sigma,r,q) = V^C(S,t,T,K,\sigma,q,r).
 \end{equation}
 By swapping the strike with the spot price and the interest rate with the dividend yield, an American call value equals the corresponding American put.  The relationship is also valid under negative discount rates \cite{Doublecontinuation2018}. Because with Equation~\eqref{eq:pcsy} we will get two option prices from one computation,  only one function evaluation is required to compute American call and put prices. We will focus on  American put options in the following sections, and compute American call options using the put-call symmetry relation.

\begin{figure}[htp]%
\centering
\subfloat[$r>q$]{{\includegraphics[width=0.5\textwidth]{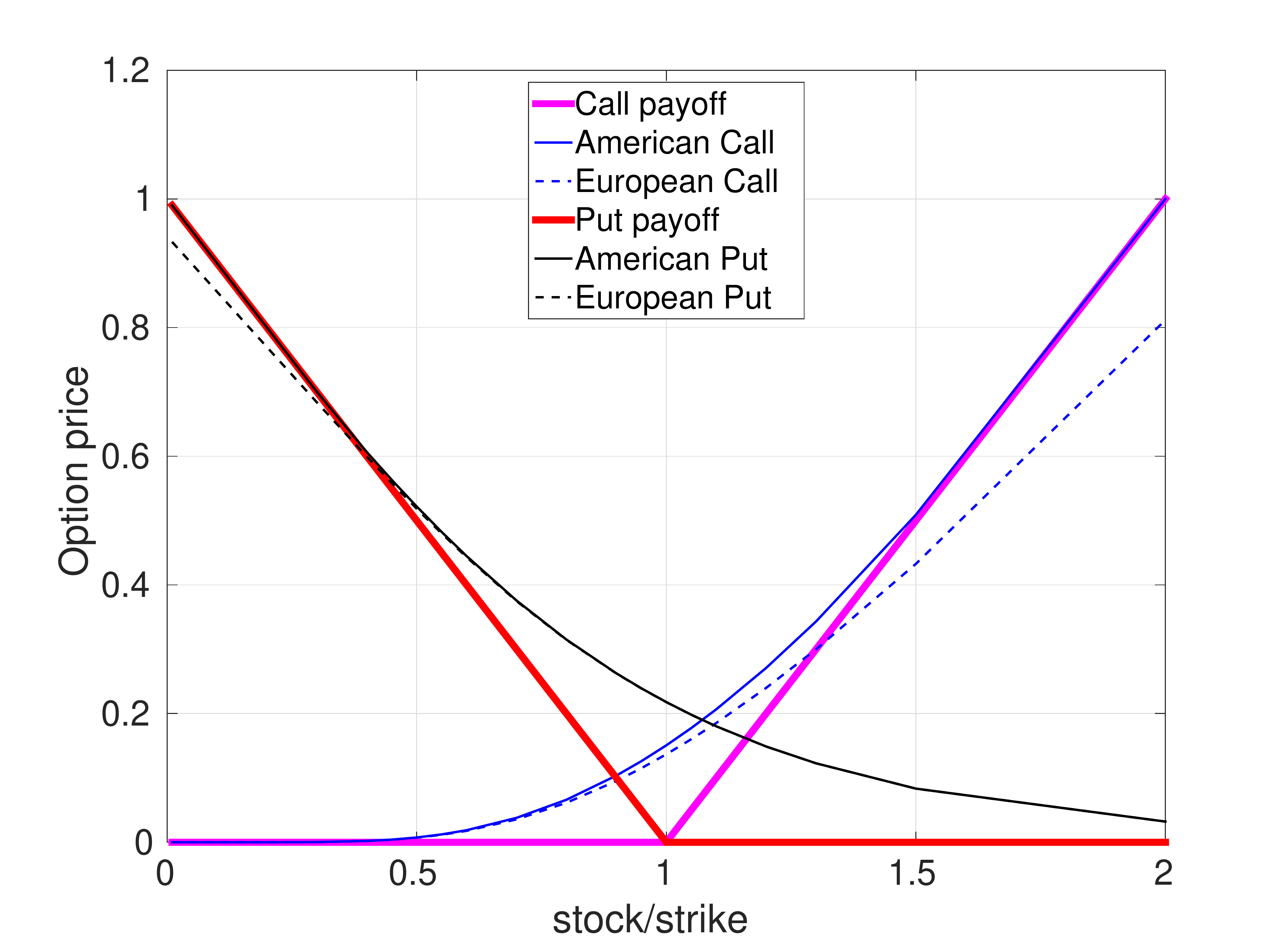} }}
\subfloat[$r<q$]{{\includegraphics[width=0.5\textwidth]{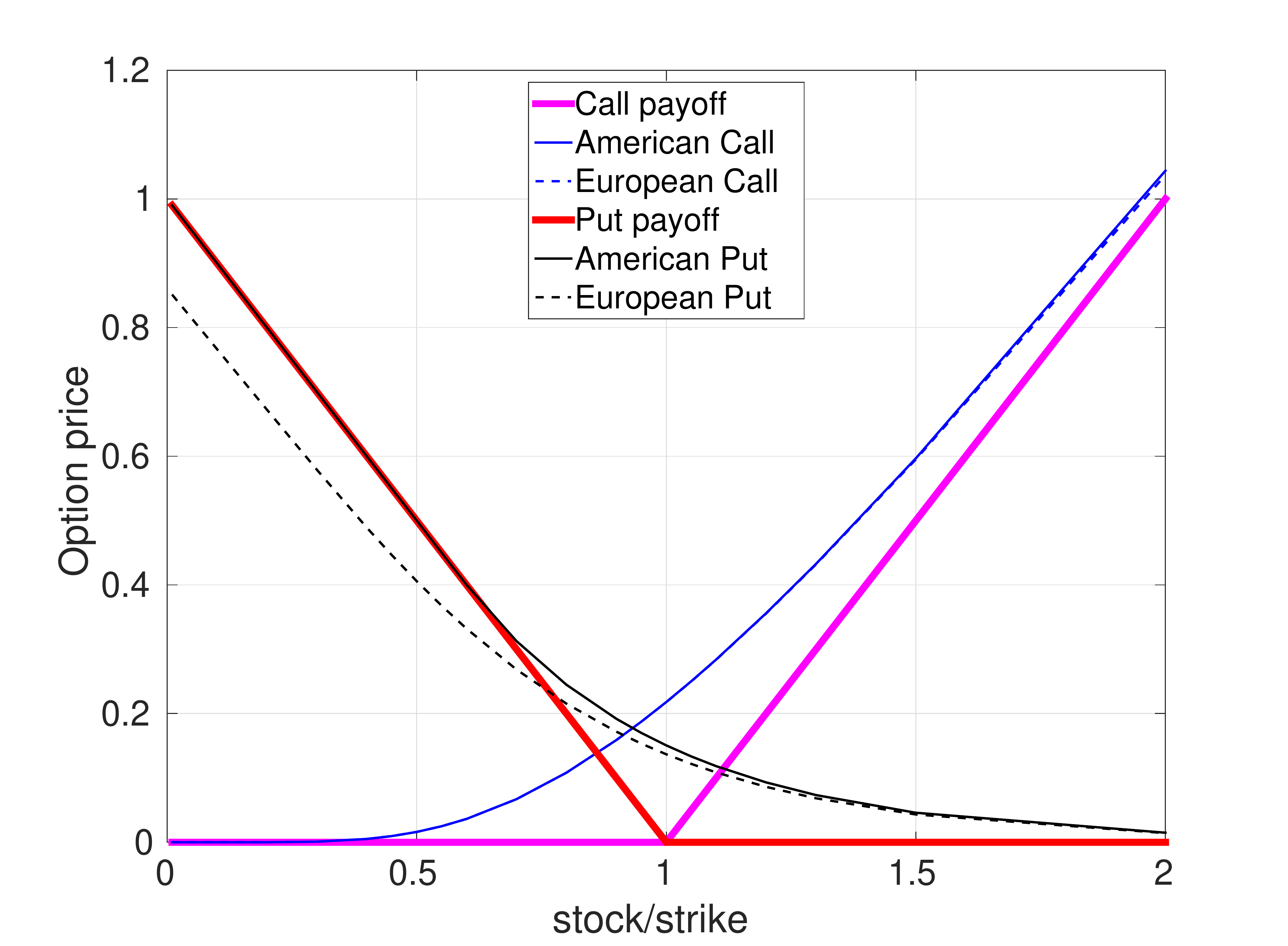}}}%
\caption{
Left: American vs. European Black-Scholes put and call option prices ($r>q$): $r=0.10$, $q=0.04$.
Right: American vs. European Black-Scholes put and call option prices ($r<q$): $r=0.04$, $q=0.10$.
The option values are  at initial time $t=0.0$, with the expiry time $T=1.5$,  volatility $\sigma=0.4$.}%
\label{fig:Am_Eu_curve}%
\end{figure}

There are two types of regions when dealing with American-style options, the continuation (or holding) and the early-exercise (or stopping) region.
The continuation region boundaries are not known a-priori. Figure~\ref{fig:Am_Eu_curve} illustrates the difference between European-style and American-style Black-Scholes option solutions, with two sets of parameters.

\begin{figure}[htp]
    \centering
    \includegraphics[width=0.7\textwidth]{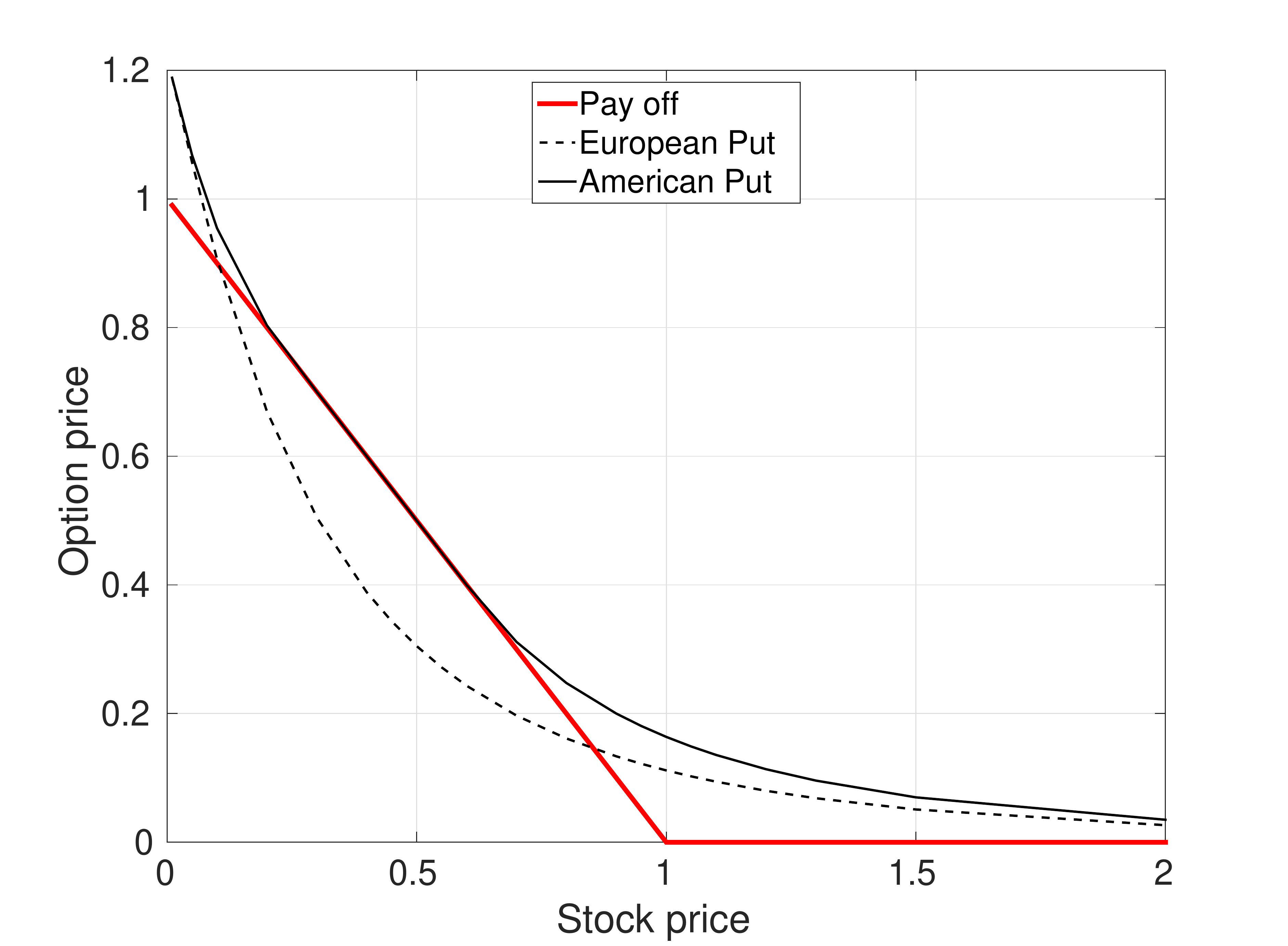}
    \caption{The setting: $r=-0.01$, $q=-0.06$, $\sigma=0.2$, $T=20$, $K=1.0$. The option value in the solid  black line  hits the payoff function twice. The stopping region is between the two early-exercise points.}
    \label{fig:Am_Eu_curve3}
\end{figure}

However, two continuation regions may arise in the case of American options, when both the interest rate and dividend yield become negative~\citep{DoubleContinuation2015,Doublecontinuation2018}. There are several reasons why we consider a negative dividend yield in our pricing model.  First of all, the interest rate may be negative in practice. When we use the put-call symmetry to compute American calls by means of puts (the counterpart), we switch the interest rate and the dividend yield in the pricing equation. For that reason, the dividend yield may also be negative in the formula.  An American-style Black-Scholes model is also used in foreign exchange or commodity markets,  where negative $q$ may be interpreted from an economic point-of-view. Figure \ref{fig:Am_Eu_curve3} presents an American put solution with two early-exercise points, so that the continuation regions are discontinuous.

%--------------------------
\subsection{Implied volatility and dividend} \label{section:American implied info}
The information implied in option prices provides  participants' expectations about future market conditions.

\subsubsection{Implied volatility}
The implied volatility of an option is the level of volatility which, inserted in the pricing model, makes the market and model prices match.
Implied volatility from market option prices is an indication for the future uncertainty of the underlying asset prices as estimated by market participants. Computing the implied volatility can be seen as an inverse problem. The American option implied volatility can be written as,
\begin{equation} \label{eq:ivf}
\sigma^* = BS_{am}^{-1}(V_{am}^{mkt};S,K,\tau,r,q, \alpha),
\end{equation}
where $BS_{am}^{-1}(\cdot)$ denotes the inversion of the American Black-Scholes pricing problem, and  $V_{am}^{mkt}$ is an American option price observed in the market.

\begin{figure}[htp]
    \centering
    \includegraphics[width=0.6\textwidth]{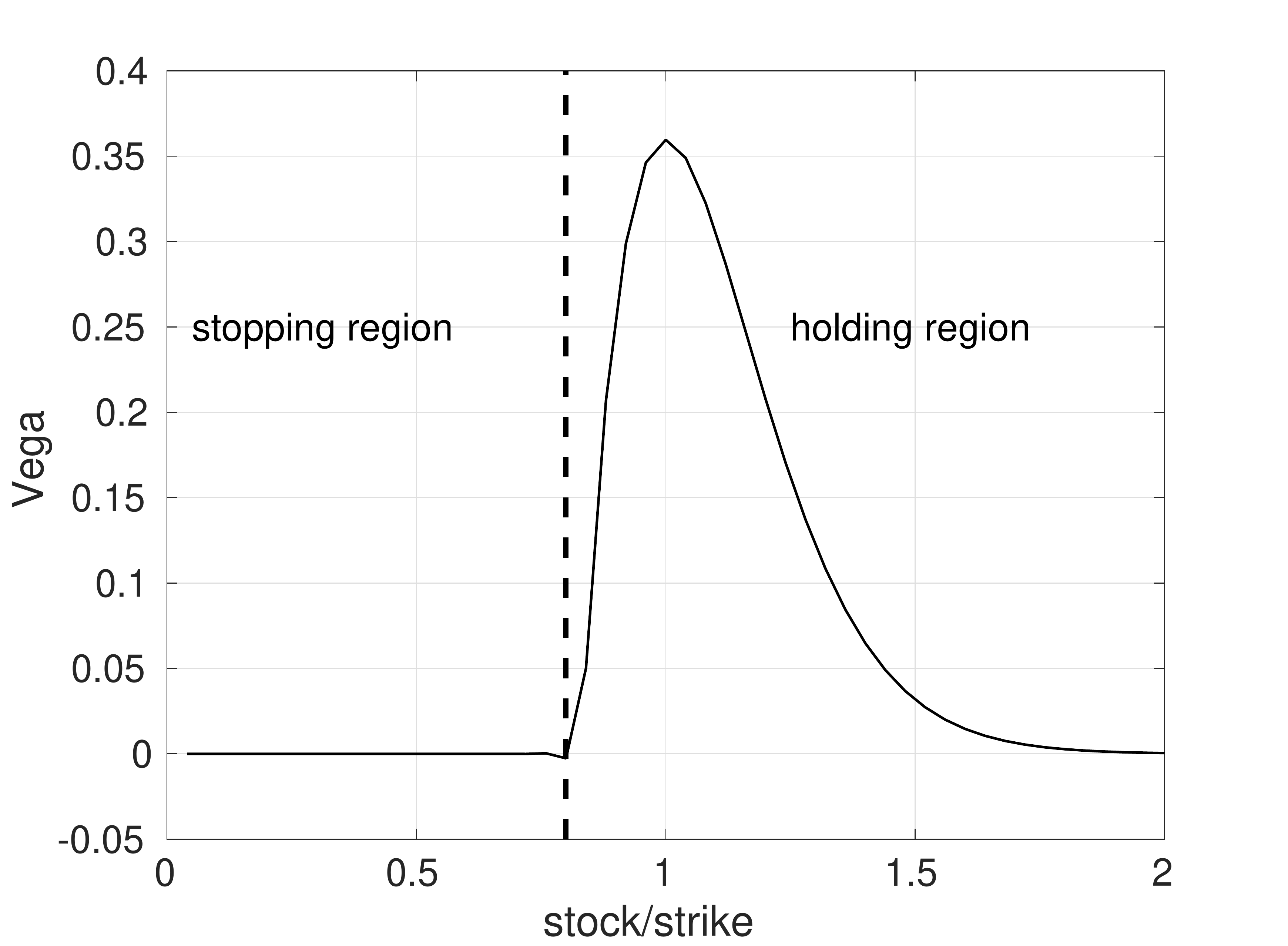}
    \caption{The Vega for American puts in different regions.}
    \label{fig:hold_stop_vega}
\end{figure}

One often solves the implied volatility problem by means of a nonlinear root-finding method, and employs an iterative algorithm to obtain its solution. Given an American option price observed in the market, the implied volatility $\sigma^*$ is determined by solving
\begin{equation}
   V_{am}^{mkt}-BS_{am}(\sigma^*; S_0,K,\tau,r,q, \alpha) =0.
\end{equation}
Existence of $\sigma^*$ can be guaranteed by the monotonicity of the Black-Scholes equation with respect to the volatility in the holding region. Unlike for European-style options, a closed-from expression for the derivative of the American option value with respect to the volatility is not available. Various solutions have been proposed to solve the implied volatility of American options, see, for example, \citep{achdou2004,CalbrationAm2019,OsherCalibrating1997,1998AmVolQam}.
As stated in~\cite{1998AmVolQam}, these solutions may have difficulties especially with deep in-the-money options. One of the reasons is that option prices are insensitive to the underlying volatility deep in the money. Gradient-free methods, like bisection, do not rely on gradient information, but they may converge slowly because of the stopping regions.
An important aspect when extracting the implied volatility is that the derivative of the option price with respect to the volatility, the option's Vega, becomes zero in the stopping region for American call  and put options.  It is well-known that,
\begin{equation} \label{eq:AmGreeks}
    |\Delta| = |\frac{\partial V_{am}}{\partial S}|=1, \;\;\;\text{Vega} = \frac{\partial V_{am}}{\partial \sigma} =0.
\end{equation}
In other words, the American option prices do not depend on the volatility in the stopping regions. As shown in Figure \ref{fig:hold_stop_vega}, Vega is positive in the holding region and zero in the stopping region. Consequently,
$$\frac{\partial \sigma}{\partial V_{am}}=\frac{1}{\text{Vega}}\rightarrow \infty.$$ When we invert the American Black-Scholes pricing problem in the stopping regions,  there is no unique solution for the implied volatility. Therefore, the definition domain of Formula \eqref{eq:ivf} should be the continuation region.

\begin{remark}
When gradient-based minimization algorithms (e.g. Newton's method) are used to extract implied information, an initial guess for the solution has to be specified. An inappropriate starting point may cause the algorithm to fall into a ``different'' continuation region from the ``envisioned'' region. Special rules have to be designed to help the algorithm reach the ``correct'' continuation region and explore the solution space. This makes it challenging to define a suitable starting point for the minimization algorithm when inverting the pricing model.
\end{remark}

In our approach, we will use the ANNs to approximate the inverse function of Formula~\eqref{eq:ivf} on the continuation region, without relying on any iterative technique, which will be explained in Section \ref{section: iv-ann-method}.

\subsubsection{Implied dividend}
Many companies pay a share of the stock value to the share holder on the ex-dividend date, which causes the stock price to drop. The option prices are also impacted by the changes in the underlying stock price.  Generally, option prices may rise (in the case of the put) or drop (the call) slightly due to the dividend payment. This dividend is called the actual dividend, denoted by $\delta$.
{\em Implied dividend} reflects how the market anticipates future dividend payments of stocks. It is extracted from option prices, and thus a quantity under the risk-neutral measure.

The difference between actual dividends and implied dividends is similar to that between historical and implied volatility. The two parameters reflect different market aspects. An implied dividend may be modeled by means of multiple components, for example,
\begin{equation}
    q = \epsilon_r + \delta + b,
\end{equation}
see~\citep{divfactors}, where $\epsilon_r$ reflects the difference between the employed and the market interest rate, $\delta$ the historical dividend and $b$ the borrowing costs of the underlying asset.
Some companies do not pay dividends, but the corresponding options may imply a non-zero dividend, which may reflect the borrowing level of the stock, see \cite{orat, J.Hull}.
The borrowing costs are seen as a factor that influences the implied dividend as a function of the time or the strike price.

Our approach is to estimate  implied dividend and implied volatility at the same time assuming the implied dividend is not constant over strike prices~\citep{amdiv2013}.
In the case of European stock options, the implied dividend can be estimated by the put-call {\em parity relation}~\cite{J.Hull},
\begin{equation}  \label{eq:putcall}
    V_{eu}^C(S,t)-V_{eu}^P(S,t) = S_te^{-q\tau}-Ke^{-r\tau},
\end{equation}
so that,
\begin{equation} \label{eq:impdiv-eu}
    q = -\frac{1}{\tau} \log(\frac{V_{eu}^C - V_{eu}^P + Ke^{-r\tau}}{S_t}).
\end{equation}
For American-style options, the put-call parity does not hold. In certain works \cite{J.Hull,CutsFromimpliedDiv2017} the authors employ Formula~\eqref{eq:impdiv-eu} to roughly estimate the implied dividend yield for American options.  Obviously, this may result in inaccuracies when the put-call parity deviation gets large.

In order to eliminate this error, the authors~\cite{phd2017} take the early-exercise premium into account for American-style options. For example, $V^C_{am}=V^C_{eu} +\Delta V^C$ and $V^P_{am} = V^P_{eu} +\Delta V^P$. Given the early-exercise premiums $\Delta V^C$ and $\Delta V^P$, Equation~\eqref{eq:putcall} can be used to calculate the implied dividend yield from the American option prices.  A requirement is that the American and European options are available, under the same parameter set. It is however usually not easily possible  to deduce the early-exercise premiums from market option prices.

Next we will numerically investigate how early-exercise premiums affect the put-call parity. As mentioned, the American option price  can be viewed as the sum of two components, the corresponding European option price and the early-exercise premium. Let $f(S_u;S_t) $  be the transition density function of  $S_u$ conditional on $S_t$ for $u\geq t$. Then, Equation \eqref{eq:AmPutIntegral} can be rewritten as follows~\cite{FinancialDerivatives},
\begin{equation}\label{eq:AmPutIntegral2}
\begin{split}
    V^P_{am}(S_t, t) = &  e^{-r(T-t)}  \int_0^K (K-S_T) f(S_T;S_t)dS_T  \\  & +  \int_t^T e^{-r(u-t)} \int_{\Omega_s} (rK-qS_u) f(S_u;S_t) dS_u du
    \\ & = V^P_{eu}(S_t, t) + \Delta V^P,
\end{split}
\end{equation}
with constant $r$ and $q$, and where $\Omega_s$ is the stopping region. If the holder of an American put chooses to exercise the put option in the case of $S_u$ being in the stopping region, he/she would gain interest $rKdt$ from the cash received, and lose dividend $qS_tdt$ from selling the asset.
Similarly, the American call price is made up of two components,
\begin{equation}\label{eq:AmCallIntegral2}
\begin{split}
     V^C_{am}(S_t, t) = & e^{-r(T-t)} \int_K^\infty (S_T-K) f(S_T;S_t)dS_T
     \\ &+ \int_t^T e^{-r(u-t)} \int_{\Omega_s} (qS_u -rK) f(S_u;S_t) dS_u du
     \\ & = V^C_{eu}(S_t,t) +\Delta V^C.
\end{split}
\end{equation}

Equations \eqref{eq:AmPutIntegral2} and \eqref{eq:AmCallIntegral2} can be substituted into the European put-call parity,
$$ (V^C_{am}-\Delta V^C) - (V^P_{am} -\Delta V^P) = S_te^{-q \tau} - Ke^{-r\tau}, $$
and the deviation from the  put-call parity is found to be,
\begin{equation} \label{eq:americandeviation}
       \Delta V^C-\Delta V^P  =V^C_{am} - V^P_{am}- S_te^{-q\tau} + Ke^{-r\tau},
\end{equation}
where $\Delta V^C$ and $\Delta V^P$ stand for the American call and put early-exercise premiums, respectively. We can measure the ``deviation'' from the European put-call parity relation  by $\text{EED}:=\Delta V^C-\Delta V^P$.
The larger the deviation, the more the American and European implied dividend yields will differ.
For European options, $\text{EED}=0$.

\begin{figure}[htp]
\subfloat[EED over maturity time  $T$ ]{\label{fig:eep_over_time}{\includegraphics[width=0.35\textwidth]{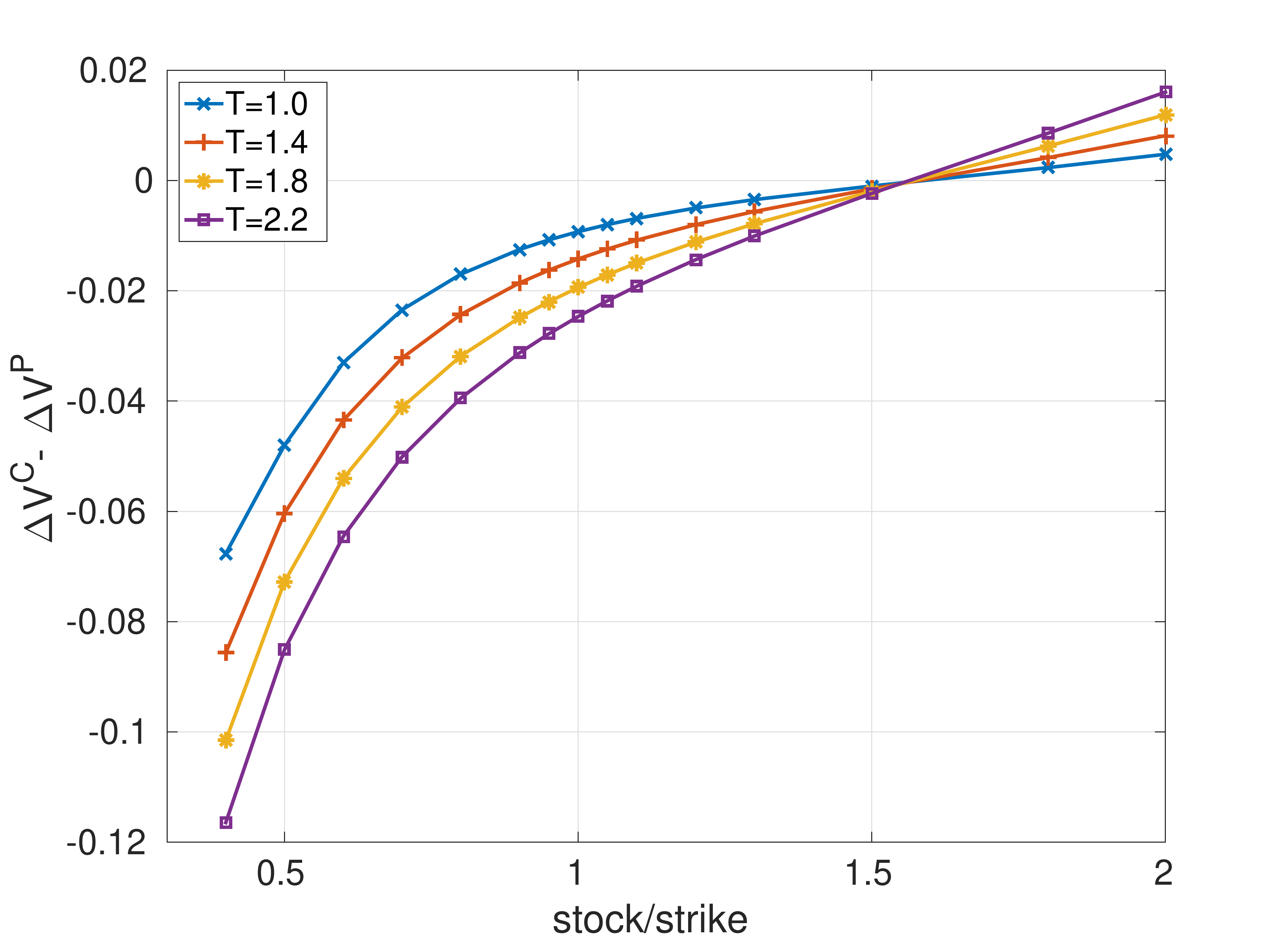}}}
\subfloat[EED against $r-q$]{\label{fig:eep_over_rq}{\includegraphics[width=0.35\textwidth]{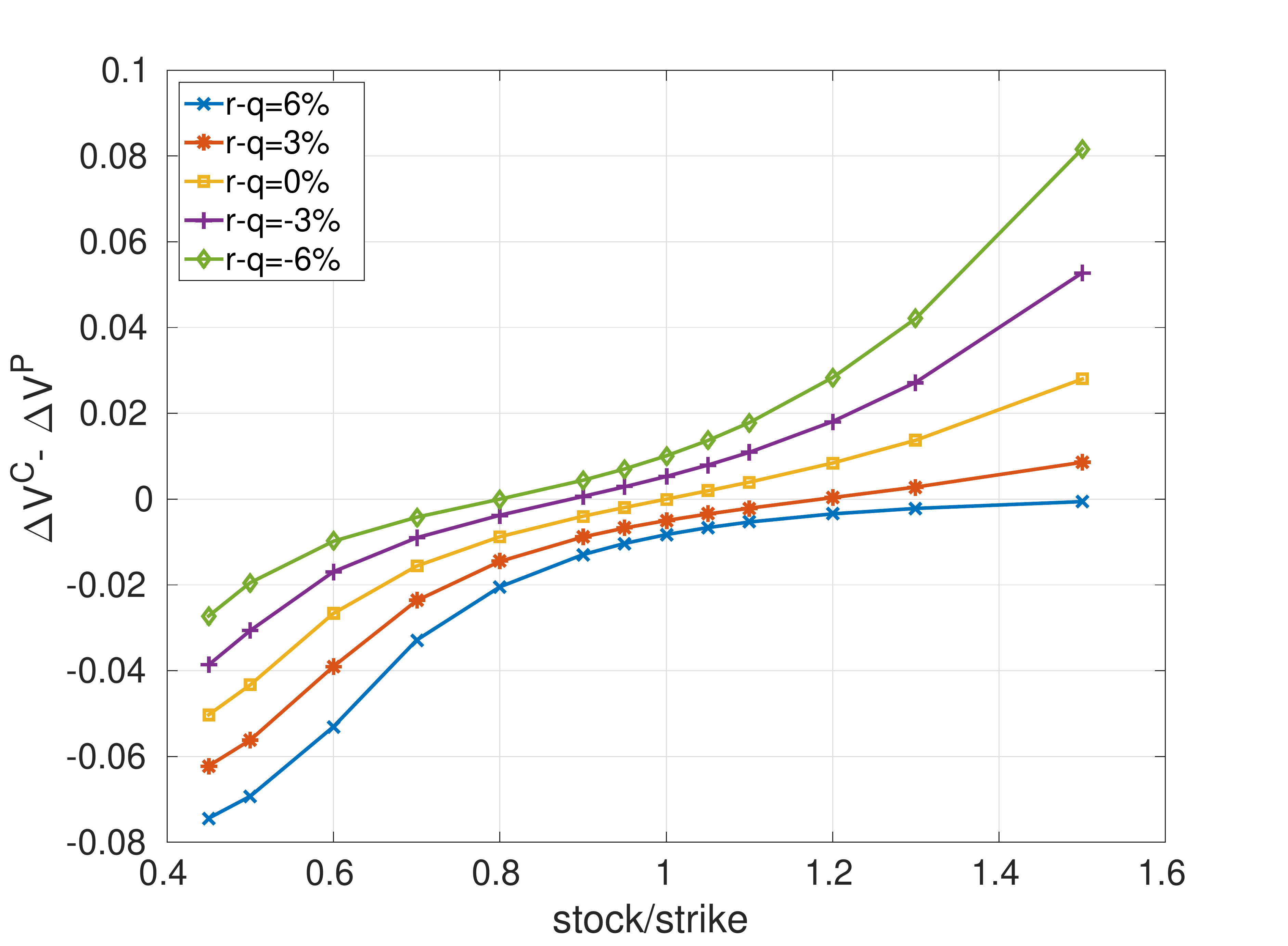}}}
\subfloat[EED against volatility]{\label{fig:eep_over_sigma}{\includegraphics[width=0.35\textwidth]{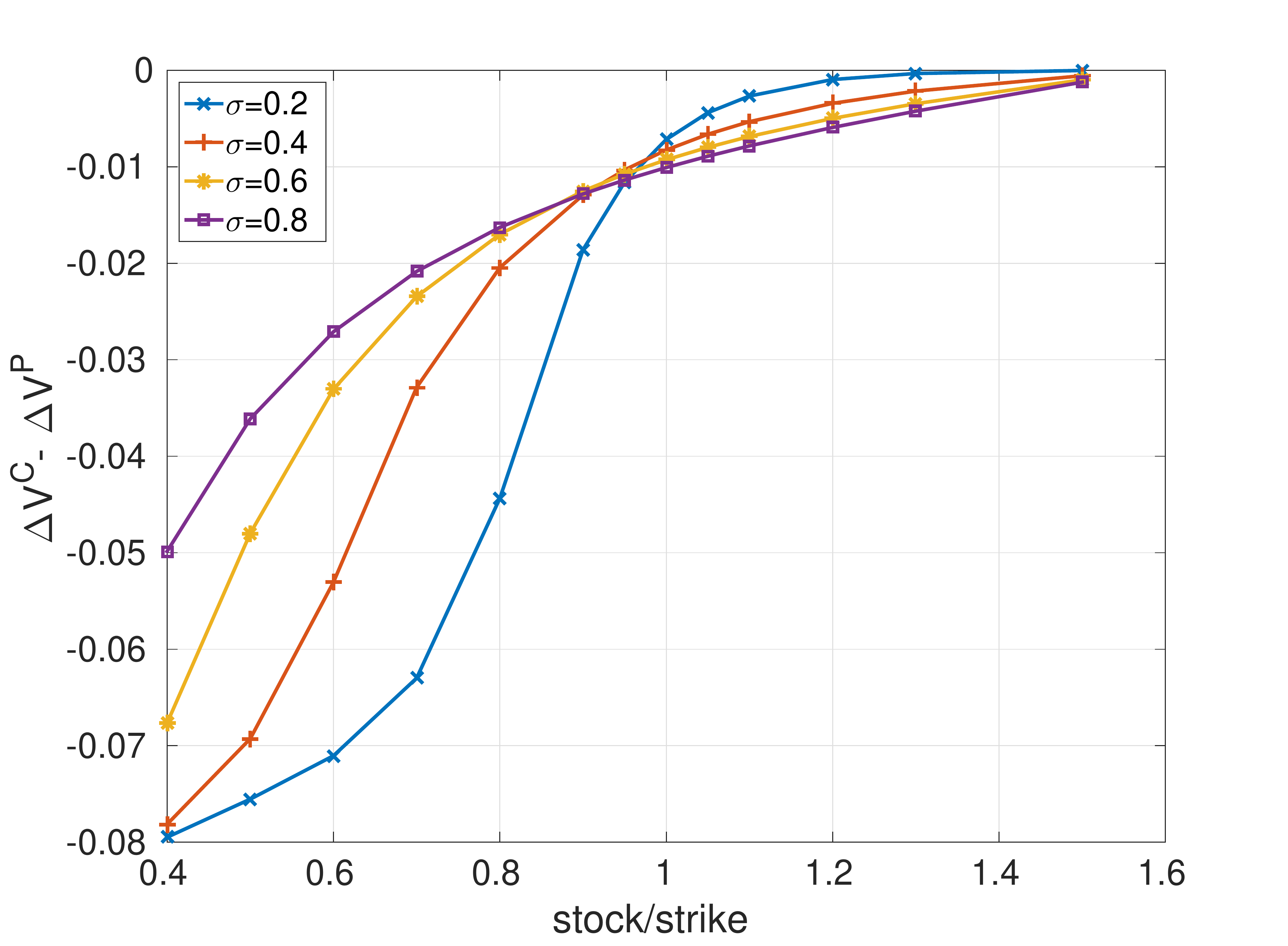}}}
\caption{Left: EED over maturity time $T$, with $\sigma=0.4$, $r=0.1$, $q=0.05$, $K=1.0$. Middle: EED against $r-q$, with $T=1.0$, $r=0.1$, $\sigma=0.4$, $K=1.0$. Right: EEP against volatility $\sigma$, with $T=1.0$, $r=0.1$, $q=0.04$, $K=1.0$.
}
\label{fig:EEPoverqandvol}
\end{figure}

We can assess the corresponding early-exercise premium by calculating the difference between the European and American option prices. In the following figures, we will demonstrate how the early-exercise premiums vary with respect to the following factors, maturity time $T$ (Fig.~\ref{fig:eep_over_time}), difference between interest rate and dividend yield $r-q$ (Fig.~\ref{fig:eep_over_rq}), volatility $\sigma$ (Fig.~\ref{fig:eep_over_sigma}). Roughly speaking, the absolute deviation is monotonically increasing when an option goes deeper into the money (out- or in-the-money, OTM or ITM), based on Figure \ref{fig:EEPoverqandvol}.  Therefore, significant errors may occur when using the European put-call parity to compute the implied dividends from American options.

%-----
\begin{remark}
The early-exercise premium (EEP) is not observable for most underlying securities, unless both American and European options with the same strike price and time to maturity are available.
Empirical studies have been conducted to analyze how the EEP varies in the market using regression techniques, like in~\cite{LI2018}.  The basic idea is to fit a function for the EEP and other observed factors, i.e.
$$ \text{EEP} = \beta_0 + \beta_1(r-q) + \beta_2(T-t) + \beta_3(S/K) +\beta_4(\sigma_{t-1}) +\epsilon $$
In the Swedish equity market, for example, the early-exercise premiums for American puts are empirically found to be positive~\cite{earlyKAmerican2010}, increasing with option moneyness, and decreasing with time to maturity and the underlying asset’s volatility.
For American-style currency options, it was observed~\cite{Geoffrey2009} that the early-exercise premiums equal approximately 5\% for puts and 4.6\% for calls.
Our numerical results coincide with these empirical studies of EEP.
\end{remark}

We will employ a model-based  approach, i.e., the American option Black-Scholes pricing model including a dividend yield is inverted in order to extract the implied dividend.
American options can be priced as follows,
\begin{equation}
  \begin{cases}
V^C_{am}=BS_{am}(\sigma^*, q^*; S_0, K, \tau, r, \alpha=1), \\
V^P_{am}=BS_{am}(\sigma^*, q^*; S_0, K, \tau, r, \alpha=-1),\\
\end{cases}
\end{equation}
where, $BS_{am}$ is the corresponding pricing model. Assuming the implied volatility and the implied dividend are the same for calls and puts with the same parameter set $K$, $S_0$, $\tau$ and $r$, there appears to be a unique solution of the system with two equations and two unknowns.  In theory, this is not always the case, as option prices do not depend on the volatility in the stopping regions. The system of equations is usually formulated as a minimization problem and a numerical optimization algorithm  is employed to search the solution space. Note that a local search based optimization method will most likely not converge when traversing those early-exercise regions.

\section{Methodology} \label{section:Methodology}

Artificial neural networks, ANNs, have been used to approximate the solution of European option pricing models by means of supervised learning, for instance, under the Black-Scholes and Heston models in~\cite{risks7010016}. Here, we will use the ANN to address the numerical solution of American-style options, and in particular use it for computing the implied information. For the inverse problem, when there is one parameter to calibrate, we can employ the ANN to build a mapping from the observed market data to the parameter to calibrate (as a unique mapping). When there are multiple parameters to calibrate, like the implied volatility and the implied dividend, the CaNN (Calibration Neural Network)~\cite{CaNN2019} is preferred.

\subsection{Artificial Neural Networks}
Neural networks are powerful function approximators. In a basic formulation, an ANN can be described as a composite function,
\begin{equation} \label{eq:fun-dnn}
\mathnormal{ \text{F}(\mathbf{x}|\boldsymbol{\theta}) = f^{(\ell)}(...f^{(2)}(f^{(1)}(\mathbf{x};\boldsymbol{\theta}^{(1)});\boldsymbol{\theta}^{(2)});...\boldsymbol{\theta}^{(\ell)})},
\end{equation}
where $\mathbf{x}$ stands for the input variables, $\boldsymbol{\theta}$ for the hidden parameters (i.e. the weights and the biases in artificial neurons), $\ell$ for the total number of hidden layers, and $f^{(\ell)}(\cdot)$ represents a hidden-layer function. The composite function, $\text{F}(\cdot)$, depends on these hidden parameters and activation/transfer functions. Once the structure, i.e. the hidden parameters and transfer functions, are determined, the ANN in Equation~\eqref{eq:fun-dnn} becomes a deterministic function.

A popular approach for training neural networks is to employ first-order optimization algorithms to determine the values of the hidden parameters which will minimize the loss function. Gradient-based algorithms are often fast, but it may be difficult to calculate the gradients for a large test set.  Stochastic gradient descent algorithms (SGD) randomly select a portion of the data set, to compute the gradient over, to address the memory issue. SGD and its variants (like Adam) are thus preferable to train the ANNs on big data sets. In a data-driven supervised learning context, the objective function is given as follows,
\begin{equation} \label{eq:argmin_dnn}
\arg \min_{\boldsymbol{\theta}} L(\boldsymbol{\theta} | (\mathbf{X},\mathbf{Y})),
\end{equation}
given the input-output pairs $(\mathbf{X},\mathbf{Y})$ and a user-defined loss function $L(\boldsymbol{\theta})$.  Thus, the ANN will be trained to approximate the function of interest in a certain norm, e.g. the $l^2$-norm. Next, we will discuss how to use the ANNs are to approximate the inverse or pricing functions.

\subsection{ANN for implied volatility} \label{section: iv-ann-method}

When we focus solely on extracting the implied volatility, the basic technique is to employ the ANN to approximate the inverse function of the American-style Black-Scholes model on a suitable effective definition domain $\Omega_h$,
\begin{equation} \label{eq:nn-bs-iv}
\sigma^* = BS_{am}^{-1}(V_{am}^{mkt};S,K,\tau,r,q, \alpha) \approx\text{NN}(V_{am}^{mkt};S,K,\tau,r,q, \alpha), \;\; [V, S,K,\tau,r,q] \in \Omega_h.
\end{equation}
The ANN is trained based on the known market variables to approximate the unique target variable $\sigma^*$.

\subsubsection{Definition domain selection}
\begin{figure}[htp]
    \centering
    \includegraphics[width=0.7\textwidth]{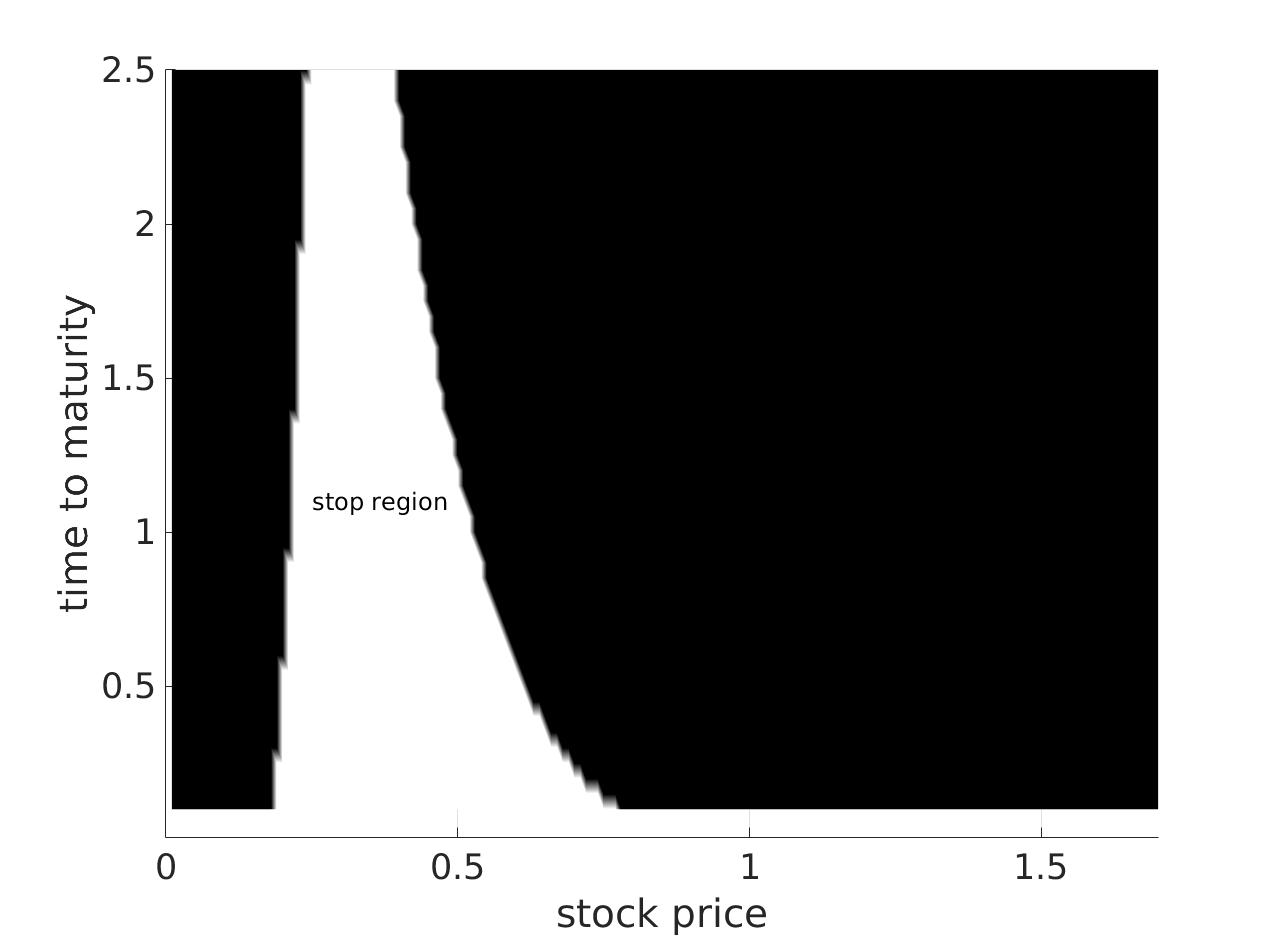}
    \caption{
Schematic diagram: An example of two continuation regions for an  American put. The shaded area represents the holding region, while the white area represents the stopping region. There are two isolated continuation regions. Here the strike price is fixed $K=1$.}
    \label{fig:am_hold_region}
\end{figure}

The effective definition domain $\Omega_h$ in Equation \eqref{eq:nn-bs-iv} corresponds to the continuation regions,  as the option value does not depend on the volatility in the stopping region for American options. The continuation regions are not known initially or are so complicated that there is no analytic formula to describe them. However,  the counterpart, the early-exercise regions,  can be found implicitly in a data-driven approach.  In our method we wish to only train the neural network on the points in the continuation region.
Overall, our aim is to find the inverse function of the American-style pricing model on the continuation region, represented by the shaded domain in Figure \ref{fig:am_hold_region}.  There is an off-line phase, where the continuation regions are being determined, and an on-line phase where we use the problem parameters to compute the implied volatility.
We build the mapping function via the ANN in the ``irregular'' continuation region.  The resulting trained ANN solver is typically much faster to determine the implied volatility than an iterative numerical solver.

First of all,  random parameter values are generated as ANN samples in the entire input domain $\Omega$, followed by detecting the parameter samples that are in the early-exercise region $\Omega_{s}$ according to Equation~\eqref{eq:freeboundary}.
We will use a robust version of the COS method~\cite{Fang2009} to calculate American option values while generating the data set during the off-line ANN phase. The COS method is an efficient numerical technique and provides the derivative information (i.e. the Greeks) essentially without any additional costs. The basic idea in~\cite{Fang2009} is to approximate the American option value by solving a small series of Bermudan-style options with different numbers of exercises opportunities, and subsequently apply a four-point Richardson extrapolation based on these four option prices. With a variety of asset and option parameters, we need to make sure that the option pricing technique is robust, i.e., it should work under all occurring parameter sets. This might imply a large integration domain within the COS method, and a relatively large number of Fourier terms in the cosine expansion.

There are two indicators to detect the samples in the early-exercise region, the difference between the option value and the payoff, and the option's sensitivity Vega in Equation \eqref{eq:AmGreeks}. We obtain the approximate continuation region, $\Omega_h = \Omega - \Omega_{s}$,  as shown in Figure~\ref{fig:am_hold_region}. The procedure requires additional computations, but these happen in the off-line phase without affecting the on-line approximation.

To control numerical errors, threshold values are prescribed for the two indicators. A threshold $\epsilon_1$ is set for the difference between the payoff function value and the generated option value,
\begin{equation} \label{eq:criterionA}
    | V_{am} (S_t,K, \tau, r, q, \sigma) - H(K,S_t) | > \epsilon_1.
\end{equation}
We also set a threshold $\epsilon_2$  for the value of Vega,
\begin{equation}\label{eq:criterionB}
    \text{Vega} >\epsilon_2.
\end{equation}
As early-exercise takes place with options that are ITM, the above two criteria only apply to ITM samples. In principle, Criterion~\eqref{eq:criterionA} should cover Criterion~\eqref{eq:criterionB}, but for robustness reasons, both will be enforced.

\subsubsection{Gradient-squashing of the option prices}
Generally, ANNs are not accurate when functions with steep gradients need to be approximated. Therefore, we need to adapt the requested output function.
In order to obtain the implied volatility from the option prices, we need to employ the gradient-squashing technique as proposed in~\cite{risks7010016}.
We subtract the intrinsic value from the American option price to obtain the corresponding {\em time value}. A brief derivation how to compute the intrinsic value of an American option with the dividend yield is given. Taking the put as an example,  recall the put-call parity for European options,
$$V^C_{eu} -S_t e^{-q\tau} =V^P_{eu} - Ke^{-r\tau}. $$
The following lower bound can be deduced,
\begin{equation}
   V^P_{eu}(S_t, t) = V^C_{eu}(S_t, t) + Ke^{-r\tau} -S_t e^{-q\tau} \geq  Ke^{-r\tau} -S_t e^{-q\tau},
\end{equation}
where the right-hand side is called the European option's intrinsic value. As an American option is at least as expensive as its European counterpart,  we have
\begin{equation}
   V^P_{am}(S_t,t) \geq V^P_{eu}(S_t, t) \geq Ke^{-r\tau} -S_te^{-q\tau}.
\end{equation}
Additionally,  American option prices should not be worth less than the pay-off function at any time, as for example shown in Figure \ref{fig:Am_Eu_curve3}, and the time value of an American put is computed by
\begin{equation}
\hat{V}^P_{am} = V^P_{am}(S_t,t) - \max(K-S_t, Ke^{-r\tau} -S_t e^{-q\tau}, 0).
\end{equation}

The gradient squashing technique~\cite{risks7010016} for computing implied volatility using the ANNs, means taking the logarithm of the time value to obtain $\log{(\hat{V}^P_{am})}$, as a quantity that can be well approximated with ANNs because its gradient is not too steep to approximate accurately.

\subsection{Determining implied dividend and implied volatility}

When the American option implied dividend yield is unknown,  we will determine both implied volatility and implied dividend simultaneously by means of the CaNN calibration methodology.  We assume the implied volatility and  the implied dividend are identical for American calls and puts with the same $K$, $S_0$, $T$, $t$ and $r$ values,
\begin{equation}
\begin{cases}
V^{C,mkt}_{am}-BS_{am}(\sigma^*, q^*; S_0, K, \tau, r, \alpha=1)  =0, \\
V^{P,mkt}_{am} - BS_{am}(\sigma^*, q^*; S_0, K, \tau, r, \alpha=-1) =0,
\end{cases}
\end{equation}
so that there are two unknown parameters to calibrate, implied volatility $\sigma^*$ and the implied dividend yield $q^*$,  given a pair of American option prices, $V^{C,mkt}_{am}$ and $V^{P,mkt}_{am}$.  The above system is reformulated as a minimization problem,
\begin{equation} \label{eq:bs-min}
    \arg \min_{ \sigma^* \in R^+, q^*\in R} (BS_{am}(\sigma^*, q^*;\alpha=1) -V^{C,mkt}_{am})^2 +(BS_{am}(\sigma^*, q^*;\alpha=-1)-V^{P,mkt}_{am})^2.
\end{equation}
As mentioned, the challenges include the relatively expensive American option pricing method, and the search technique within the optimization where convergence
may hamper when the iterative optimization method enters the region where Vega=0.

We adapt a fast, generic and robust calibration framework, the CaNN (Calibration Neural Networks) developed in~\cite{CaNN2019}.  The basic idea of the methodology is to convert the calibration of model parameters into an estimation of a neural network’s hidden units. The reason for this is that model calibration and training ANNs
(here supervised learning) can be reduced to solving an optimization problem according to the Equations~\ref{eq:argmin_dnn} and~\ref{eq:bs-min}. It enables parallel GPU computing to speed up the computations, which makes feasible to employ a global optimization technique to search the solution space.  The employed gradient-free optimization algorithm, Differential Evolution (DE),  does not get stuck in  local minima or in the stopping region.  Another benefit of DE is that it is an inherently parallel technique where populations of possible optimal solutions can be evaluated simultaneously within the ANN.

\begin{figure}[htp]%
\centering
\subfloat[Training phase]{\label{fig:training phase} {\includegraphics[width=0.45\textwidth]{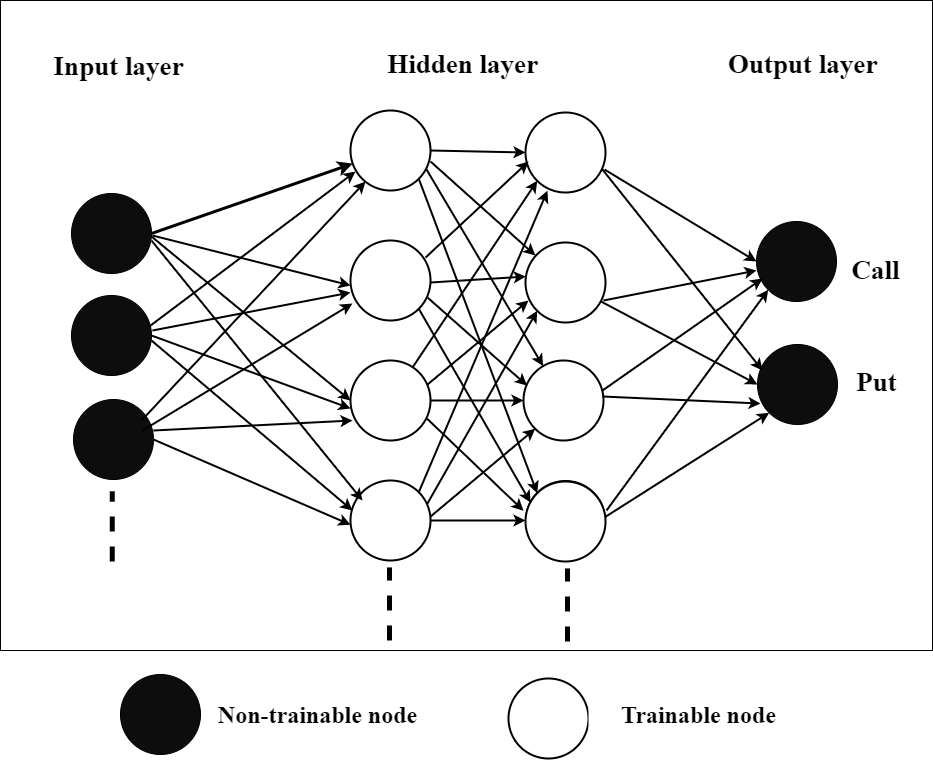} }}%
\qquad
\subfloat[Calibration phase ]{\label{fig:calibrating phase}{\includegraphics[width=0.45\textwidth]{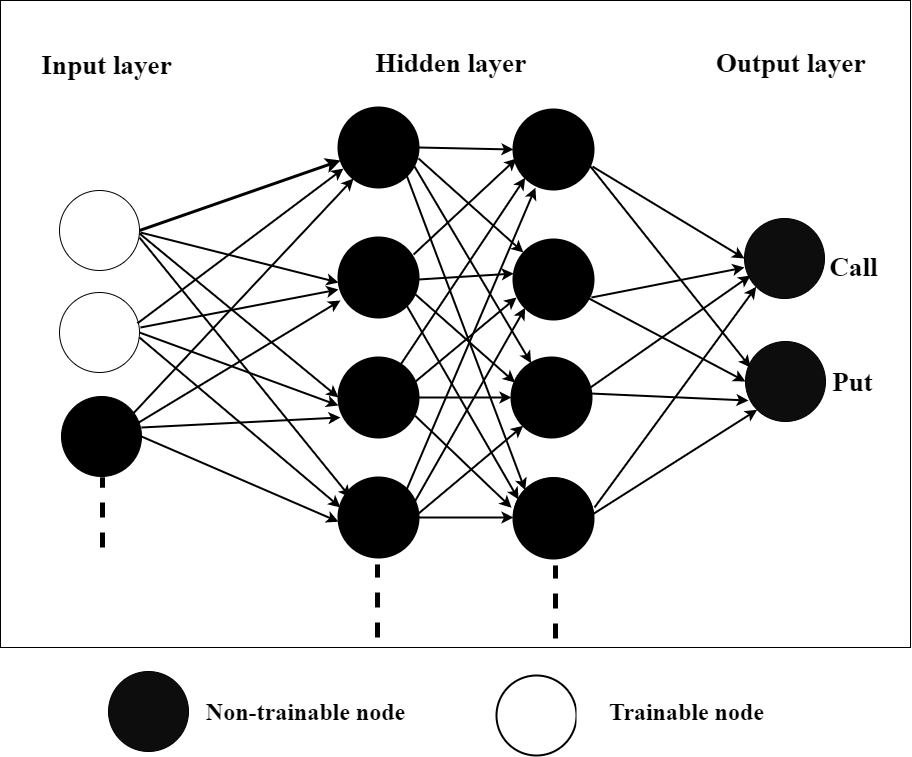}}}%
\caption{ Left: In the forward pass of the CaNN, the output layer produces two option prices. Right: In the calibration phase, the CaNN estimates the two parameters, implied volatility and implied dividend, in the original input layer.}
\label{fig:two_output_CaNN}%
\end{figure}

There is a calibration phase (the backward stage) in the CaNN, in addition to the training and testing phases.  The training phase is to determine suitable weights and biases in the hidden layers to map model input to output, while the calibration phase estimates the model input parameters to optimally match
a given output.  We will view these three phases as a whole framework, and just change the learnable units between the original layers (i.e. the hidden, output and input layers) within the different phases.
More specifically, the CaNN consists of two stages.
The forward pass, including the training and testing phase, approximates the American-style Black-Scholes prices.
We have developed one neural network providing two output values in the forward pass, the American call and the put prices, as illustrated in Figure~\ref{fig:two_output_CaNN}.
The backward pass, on the other hand,  aims to find the two parameters, $(\sigma^*,q^*)$, to match the two observed American option prices, $V^{P,mkt}_{am}$ and $V^{C,mkt}_{am}$, with strike price $K$, maturity time $T$, spot price $S_0$, interest rate $r$.

This is different from the network in~\cite{CaNN2019}, where one neural network corresponds to one output quantity.  Therefore Equation \eqref{eq:bs-min} is written as an objective function of model calibration,
\begin{equation} \label{eq:nn-min}
\arg \min_{\sigma^* \in R^+, q^*\in R} ( \text{NN}(\sigma^*, q^*;\alpha=1) -V^{C,mkt}_{am})^2 +(\text{NN}(\sigma^*, q^*;\alpha=-1)-V^{P,mkt}_{am})^2,
\end{equation}
which is used as the loss function for the backward pass in the CaNN.

\begin{remark}
Because of a generic calibration framework, the CaNN can easily deal with more complex situations, for example, the objective function~\eqref{eq:nn-min} includes more than a pair of American price quotes which share the same implied dividend, for example, as in the work~\cite{Qi2005}.
\end{remark}

\subsection{The ANN architecture}
Our chosen ANN architecture in the forward pass constitutes four hidden layers and two parallel output layers. Some particularly useful operations in deep neural networks, e.g. dropout and batch normalization, do not bring any significant benefits in our ``shallow'' ANN. The proposed configuration has been demonstrated to be able to fit the pricing model with acceptable accuracy in~\cite{CaNN2019}. The activation function Softplus is chosen in this paper, i.e.
$$\hat{f}(x) = \log(1+e^x),$$
as its smooth derivative fits well to the smoothness of the pricing function, especially in the continuation region.  According to the universal approximation theorem, a one-layer based ANN can be used to approximate any continuous function to any desired precision, but with the rate which linearly depends on the number of neurons involved. The depth of the ANN (i.e. the number of hidden layers) can increase the function's representation accuracy exponentially, but deep ANNs becomes difficult to implement in parallel (e.g. the current hidden layer has to wait for output signals of the previous one), resulting in long computation times. Considering the approximation power and the computation efficiency, we choose four hidden layers and 200 neurons in each layer. Both ANNs in the forward and backward stages of the CaNN will make use of the hyper-parameters that are shown in Table \ref{table:ann-setting}.  In the backward stage, the DE algorithm, a gradient-free global optimizer, replaces Adam. These values have shown to result in a robust neural network which converges well for a variety of problem parameters.

 \begin{table}[htp]
\begin{center}
\caption{The ANN configuration. }
 \begin{tabular}{ c | c }
  \hline
  Hyper-parameters     & Options \\ \hline
  Hidden layers & 4 \\
  \hline
  Neurons (each layer) & 200 \\
  Activation     & Softplus \\
  Initialization   & Glorot\_uniform\\
  Optimizer      & Adam \\
  Batch size     & 1024 \\
  \hline
 \end{tabular}
 \label{table:ann-setting}
 \end{center}
 \end{table}

%---------------------------------------
\section{Numerical results} \label{section:numerical results}

This section presents numerical experiments for using the ANN to extract implied information from American options. We use the COS method \citep{Fang2009}  to compute American option prices for the training data set, see Appendix~\ref{sec:am-cos}. The basic procedure includes computing Bermudan options using COS under the Black-Scholes framework, followed by employing a four-point Richardson extrapolation to compute the price of the American options~\cite{Fang2009}.

\subsection{Implied volatility}

Without loss of generality, we use a fixed spot price $S_0=1.0$. Then, the input for the ANN  is made up of five parameters $\{\log{(\hat{V}^P_{am})}, K, r,q, \tau\}$. The two thresholds in the Equations~\eqref{eq:criterionA} and~\eqref{eq:criterionB} are $\epsilon_1 = 0.0001$ and $\epsilon_2 = 0.001$ for the data set.
 \begin{table}[htp]
\begin{center}
 \caption{ Train dataset for American options under the Black-Scholes model; The spot price $S_0=1$ is fixed. The upper bound of American put price is 1.2. LHS stands for Latin Hypercube Sampling. }
 \begin{tabular}{ c|c | c | c }
  \hline

  ANN & Parameters       & Value range   & Employed method \\ \hline

  \multirow{4}{*}{ANN Input} & Strike, $K$     & [0.6, 1.4] & LHS \\
  & Time value, $\log{(\hat{V}^P_{am})}$        & $(-11.51, -0.24)$ & COS\\

  &Time to maturity, $\tau$    & [0.05, 3.0] & LHS \\

  &Interest rate, $r$     & [-0.05, 0.1] & LHS \\
  &Dividend yield, $q$     & [-0.05, 0.1] & LHS \\
  \hline
  ANN output& Implied volatility, $\sigma^*$        & $(0.01, 1.05)$ & LHS\\

  \hline

 \end{tabular}
 \label{table:amput_iv_ann}
 \end{center}
 \end{table}

The ANN is trained with American put options to learn the weights of the ANN-based solver for the computation of the implied volatility from American options.
We define the measures as follows:
$$\text{MSE} = \frac{1}{n} \sum_{i=1}^n (y_i-\hat{y}_i)^2,\;\;\; \text{MAE} = \frac{1}{n} \sum_{i=1}^n |y_i-\hat{y}_i|, \;\;\;\text{MAPE} = \frac{1}{n} \sum_{i=1}^n \frac{|y_i-\hat{y}_i|}{y_i},$$
where $y$ represents the true value of American option prices, and $\hat{y}$ represents the predicted value,  with $n$ being the number of samples. During training,  $\text{MSE}$  is used to find the weights and biases,  while the other two measures $\text{RMSE}$ and $\text{MAPE}$ are monitored. The goodness of fit, $\text{R}^2$,  is also provided, which describes the closeness between the predicted values and the true values. A single measure may not be reliable, while multiple measures can provide different angles to evaluate the performance.

After the model input parameters are sampling (here by LHS) over the specific domain, the COS method is used to solve their corresponding American-style Black-Scholes pricing model, reaching the data collection $\{S_0,K,\tau,r,q,\sigma,V_{am}\}$. Afterwards, the variable $\sigma$ is placed into the output layer of the ANN, as the implied volatility $\sigma^* \equiv \sigma$ for the data collection. Meanwhile, the other variables in the collection are involved in the input layer of the ANN, and more details are in Table~\ref{table:amput_iv_ann}. The validation samples help avoid over-fitting training the ANN. The test samples which the ANN did not encounter during training are subsequently used to evaluate the generalization performance of the trained ANN.  There are around one million samples, with 80\% as training, 10\% as validation, 10\% as test phase samples.  The learning rate is halved every 400 epochs during training.  After 4000 epochs, the training and validation losses have converged.

\begin{figure}[htp]%
\centering
\subfloat[Training]{{\includegraphics[width=0.5\textwidth]{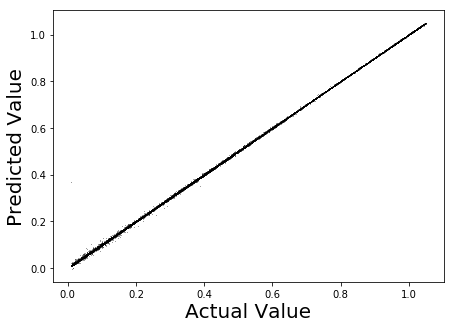}}}%
\subfloat[Testing]{{\includegraphics[width=0.5\textwidth]{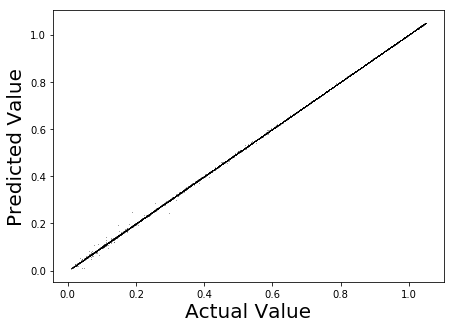} }}
\caption{Left: $\text{R}^2$=0.999994; Right: $\text{R}^2$=0.999993}%
\label{fig:AmPutR2}%
\end{figure}

Table \ref{table:perfomance_iv_AmPut} and Figure \ref{fig:AmPutR2} present the performance of the trained ANN.
The test performance is close to the train performance, suggesting that the trained ANN generalizes well for unseen data, as shown in Table \ref{table:perfomance_iv_AmPut}. The ANN predicted implied volatility values approximate the true  values  accurately  for  both  the  train  and  test  datasets,  as is indicated  by the $\text{R}^2$
measure in Figure~\ref{fig:AmPutR2}. Moreover, the online prediction phase for the American-style implied volatility, requiring only the evaluation of the trained ANN, is much faster than traditional iterative root-finding algorithms.

It is observed that the trained model performance tends to decrease when the pricing model parameters gets close to the upper or lower bounds of the values in Table \ref{table:amput_iv_ann}.  In other words, outliers are most likely to appear near the boundary. Thus the training data set is recommended to have a wider parameter range than the test range of interest.

 \begin{table}[htp]
\begin{center}
\caption{ Multiple measures are used to evaluate the performance.  }
\scalebox{1.0}{
   \begin{tabular}{  c | c | c | c | c}
    \hline
    - &  MSE & MAE & MAPE & $\text{R}^2$  \\ \hline
    Training  & 4.33 $\cdot 10^{-7}$  &  2.44$\cdot 10^{-4}$  & 1.11$\cdot 10^{-3}$  &  0.999994\\
    \hline
    Testing & 4.60$\cdot 10^{-7}$   &  2.51$\cdot 10^{-4}$   & 1.15$\cdot 10^{-3}$   & 0.999993\\
    \hline
  \end{tabular}  }
\label{table:perfomance_iv_AmPut}
\end{center}
\end{table}

\subsection{CaNN for implied information}

The CaNN is a generic framework for the fast calibration of financial models, and involves two stages, a forward pass and a backward pass. We implement the forward and backward pass in a sequential way.

Here we extend the original CaNN by using one forward pass to approximate two American option prices, a put and a call value.  The input for the neural network in the forward pass consists of $(S_0,K,r,q,\sigma,\tau)$, and the output compromises a pair of American prices, that is $(\tilde{V}^P_{am}, \tilde{V}^C_{am})$.   As the neural network gives us two output variables, the loss function of the forward pass includes two components,
\begin{equation}
    \text{MSE} = \frac{1}{2n} \sum_{i=1}^n \{ (\tilde{V}^P_{am,i}-V^{P,mod}_{am,i})^2 + (\tilde{V}^C_{am,i}-V^{C,mod}_{am,i})^2 \}
\label{mse22}
\end{equation}
where $V_{am}^{P,mod}$  and $V^{C,mod}_{am}$ stands for the American put and call prices, respectively, that are generated by the American-style Black-Scholes model. This rule also applies to MAE and MAPE.
There are four hidden layers with 200 neurons each layer, as shown in Table \ref{table:ann-setting}. The total number of hidden  parameters is 122,202, and the loss function Equation~\ref{mse22} is used to update the hidden layers of the CaNN during the training phase. The training data set is constructed according to the parameter ranges in Table \ref{table:forward ANN}. In Table \ref{table:ANN-IV-Heston Performance} the performance of the CaNN is presented. The results, for both Calls and Puts, are highly satisfactory, achieving very good levels of precision in all the considered measures.

\begin{table}[htp]
\begin{center}
 \caption{ Training data set for the forward pass. We fix $S_0=1$, and sample strike prices $K$ to generate different moneyness levels. The total number of the data samples is nearly one million, with 80\% training, 10\% validation, 10\% test samples. }
 \begin{tabular}{ c|c | c | c }
  \hline

  ANN & Parameters       & Value range   & Method \\ \hline

  \multirow{5}{*}{Forward input} & Strike, $K$     & [0.45, 1.55] & LHS \\

  &Time to maturity, $\tau$    & [0.08, 3.05] & LHS \\

  &Risk-free rate, $r$     & [-0.1, 0.25] & LHS \\

  &Dividend yield, $q$     & [-0.1, 0.25] & LHS \\
  &Implied volatility, $\sigma$        & $(0.01, 1.05)$ & LHS \\
  \hline
  \multirow{2}{*}{Forward output} & American put, $V^P_{am}$          & $(0, 1.8)$ & COS\\
   & American call, $V^C_{am}$& $(0, 1.2)$ & COS\\
   \hline

 \end{tabular}
 \label{table:forward ANN}
 \end{center}
 \end{table}

\begin{table}[htp]
\begin{center}
\caption{The performance of the CaNN forward pass with two outputs.}
\scalebox{1.0}{
 \begin{tabular}{ c | c | c | c | c |c  }
  \hline
     -- & Option & MSE & MAE & MAPE & $\text{R}^2$   \\ \hline
   \multirow{2}{*}{Training} & Call &  $1.40\times 10^{-7}$  & $ 3.00\times 10^{-4}$& $ 1.25\times 10^{-3}$ & 0.9999965 \\
   &Put &  $2.54\times 10^{-7}$  & $ 4.24\times 10^{-4}$& $ 1.64\times 10^{-3}$ & 0.9999959 \\
   \hline
   \multirow{2}{*}{Testing} & Call &  $1.43\times 10^{-7} $ & $3.02\times 10^{-4}$ & $1.27\times 10^{-3}$ & 0.9999964 \\
    &Put &  $2.55\times 10^{-7} $ & $4.26\times 10^{-4}$ & $1.64\times 10^{-3}$ & 0.9999959\\

  \hline
\end{tabular} }
\label{table:ANN-IV-Heston Performance}
\end{center}
\end{table}

After performing the forward pass, the CaNN's backward pass carries out the actual calibration. Supposing each option quote in the market includes American call and put prices, the idea behind the backward pass is to determine two parameters $(\sigma^*,q^*)$ within the American Black-Scholes model to best match the pair of market option prices, given the interest rate $r$,  maturity time $T$, strike price $K$, and spot price $S_0$.
The  objective function for the calibration procedure is found in Formula \eqref{eq:nn-min}, which is equivalent to Criterion~\ref{mse22} in the case of $n=1$.  In practice, the market price is taken to be the mid-price of the bid and ask prices.
\begin{remark}
The objective function can, of course, also be defined in a different way to take into account the bid-ask spread. This is however out of our scope here.
\end{remark}

In order to evaluate the approach, we prescribe model parameters and investigate how accurately the CaNN can recover them. Table \ref{table:AmCaNNexamples} presents a set of
examples, including many different scenarios, e.g. ITM and OTM scenario's are considered. The results in Table \ref{table:AmCaNNexamples} suggest that the CaNN can recover the implied volatility and implied dividend highly accurately from ``artificial market option data''. Even when interest rates and dividend yields are negative, the CaNN recovers the true values without stalling convergence. The method's robustness may be attributed to the robust numerical solver generating accurate option prices for a wide range of model parameters, the designed neural network providing sufficient approximation capacity, and the gradient-free optimizer (i.e. DE) to globally search the solution space.
\begin{table}[htp]
\begin{center}
\caption{Examples of using CaNN to extract implied volatility and implied dividend.  $\dag$ represents the prescribed values, $*$ represents the calibrated values.}
\begin{tabular}{c c c  c c  c c | c c  }
 \hline
$K/S_0$  &  $T$ & $r$ &  $\sigma^\dag$ & $q^\dag$ & $C_{am}^{mkt}$ & $P_{am}^{mkt}$ & $\sigma^*$ & $q^*$  \\
\hline
1.0 & 0.5 & -0.04 & 0.1 & 0.06 & 0.0146 & 0.0597 & 0.099 & 0.059 \\
1.1 & 0.5 & -0.04 & 0.2 & -0.06 & 0.0255 & 0.1181 & 0.198 & -0.061 \\
1.0 & 0.75 & 0.0 & 0.3 & -0.02 & 0.1119 & 0.0976 & 0.300 & -0.020 \\
1.2 & 1.0 & -0.04 & 0.4 & 0.08 & 0.0603 & 0.3810 & 0.40 & 0.080 \\
0.8 & 1.0 & 0.02 & 0.3 & 0.02 & 0.2322 & 0.03472 & 0.299 & 0.020 \\
0.7 & 1.25 & 0.0 & 0.4 & -0.04 & 0.3886 & 0.0378 & 0.399 & -0.040  \\
 \hline
\end{tabular}
\label{table:AmCaNNexamples}
\end{center}
\end{table}

For DE,  a  search interval  is required for each parameter, we provide [-0.08,0.1] for $q^*$ and (0, 1.0) for $\sigma^*$. During the mutation stage, we set the population size of each generation to be a small number, here 10. We choose 'best1bin' as the mutation strategy, that is, the best candidate of the previous generation enters the mutation stage and there are two items in the differential operator. The differential weight is chosen randomly between 0.5 and 1.0. During the crossover stage, the crossover possibility is 0.7. During the selection stage,  all new trial candidates with the objective function can be processed in parallel, and the convergence tolerance is set to 0.01.  As the number of calibration parameters is only two, the computation time on a CPU is only around 0.37 seconds using the sequential DE and is less than 0.1 second using the parallel version of the DE method.

Furthermore, a systemic test is conducted to assess the averaged performance  over a large number of cases. We generate equally-spaced samples over a certain interval according to Table \ref{table:systest}, but remove the samples that are connected to early-exercise region option prices. The experiment ends up with 9271 test cases.  The results in Table \ref{table:system-am-CaNN} suggest that the proposed approach performs well under a wide variety of option market conditions at a reduced computational cost.  The calibration speed is due to the efficient forward pass and the parallel, gradient-free DE optimizer.  Basically the forward pass serves as a fast numerical solver for the American pricing model. Additionally, with two output prices, the forward pass requires half of the computation cost.

\begin{table}[htp]
\begin{center}
\caption{The parameter range for the systemic experiment.}
\begin{tabular}{c | c c  c  }

 \hline
Parameter & interval & step & number  \\
\hline
$\sigma$ & [0.1, 0.45] & 0.05  & 8\\
$q$ & [-0.06, 0.08] & 0.02  & 8 \\
\hline
$K$ & [0.7, 1.2] & 0.1  & 6\\
$\tau$ & [0.5, 1.5] & 0.25  & 5\\
$r$ & [-0.04, 0.06] & 0.02  & 6\\
$P_{am}^{mkt}$ & [0.7, 1.2] & -  & 9271\\
$C_{am}^{mkt}$ & [0.7, 1.2] & -  & 9271 \\
 \hline
\end{tabular}
\label{table:systest}
\end{center}
\end{table}

 \begin{table}[htp]
 \caption{The averaged performance of the CaNN estimating implied volatility and implied dividend based on 9271 different test cases. The averaged number of  function evaluations is 1060. }
 \begin{center}\footnotesize
 \renewcommand{\arraystretch}{1.3}
 \begin{tabular}{ lr | lr}\hline\hline
 \multicolumn{2}{c}{ Absolute deviation }  & \multicolumn{2}{c}{Computational cost}\\ \hline
 $|\sigma^\dag-\sigma^*|$     & $6.65\times 10^{-3}$
  & \textbf{CPU} time (seconds) & $0.08$ \\
 $|q^\dag-q^*|$     & $8.56\times 10^{-4}$
    & \textbf{GPU} time (seconds) & $0.04$ \\
 \hline\hline
 \end{tabular}\label{table:system-am-CaNN}
 \end{center}
 \end{table}

%---------------------------------------
\section{Conclusion}
We studied the problem of pricing American-style options  and  extended a data-driven machine learning method to extract
the implied volatility and/or implied dividend yield from observed market American option prices in a fast and robust way.

For computing the American implied volatility,  we explained that the domain for the inverse function should be equivalent to the continuation regions of the American options.
The ANN-based approach builds an approximating function and addresses complex boundaries of the definition domain, by means of  the different off-line and on-line phases. More specifically, we used two conditions to  classify the random data samples in the domain in the off-line phase.   The definition domain is represented  by  data points which lie in the continuation regions.  Subsequently, a neural network was trained on those samples to approximate the inverse function.  This data-driven approach  avoids an iterative algorithm  which may suffer from convergence problems.  Due to the off-line definition of the domain, our approach also successfully dealt with negative interest rates and dividend yields, where two early-exercise regions may appear. In short, the offline-online decoupling brings much flexibility.

Furthermore, we presented a method for finding simultaneously implied dividend and implied volatility from American options using a calibration approach. The CaNN, which consists of an efficient solver and a fast global optimizer, is employed to carry out the calibration procedure. As a result, the early-exercise premiums, which the European option put-call parity relation fails to deal with, are handled successfully. The parallel global optimizer prevents the CaNN from stopping in the early-exercise regions and allows to achieve a good quality solution in a short amount of time.  The numerical experiments demonstrate that the CaNN is able to accurately extract multiple pieces of implied information from American options.  A continuous dividend yield is considered in this paper,  and it should be feasible to extend the approach to deal with time-dependent or discrete dividends.

%---------------------------------------
\section{Acknowledgements}
The authors would like to thank the China Scholarship Council (CSC) for the financial support. We  would also like to thank Dr.ir Lech Grzelak for valuable suggestions, as well as Dr. Damien Ackerer for fruitful discussions.

\bibliographystyle{unsrt}
\bibliography{main}

%\newpage
\appendix
\section{Pricing American options by the COS method}\label{sec:am-cos}
For the valuation of American options, we use Richardson extrapolation on a series of Bermudan options with an increasing number of exercise opportunities.

We start with a Bermudan option, where the holder has the right to exercise  the contract at pre-specified dates before maturity. With $t_0$ being initial time, we assume there are $M$ pre-specified exercise dates, and have $\{t_1, \cdots, t_M\}$
as the collection of all exercise dates. The regular time interval reads $\Delta t:=(t_m-t_{m-1})$, $t_0<t_1<\cdots<t_M=T$. With the help of the risk-neutral valuation, we arrive at the pricing formula for a Bermudan option with $M$ exercise dates, for $m=M,M-1, \ldots,2$:
\begin{eqnarray}
\setlength\arraycolsep{3pt}
\left\{ \begin{array}{lll}
c(x, t_{m-1}) &=& e^{-r\Delta t}\int_{\mathbb{R}}^{} V(y, t_{m}) f(y|x) dy, \label{berm1}\\[1.5ex]
V_{ber}(x, t_{m-1})&=& \max\left(h(x,t_{m-1}), c(x,t_{m-1})\right),
\end{array} \right.
\end{eqnarray}
followed by
\begin{equation} \label{v0berm}
V_{ber}(x,t_0)=e^{-r\Delta t}\int_{\mathbb{R}} V(y, t_{1}) f(y|x)dy.
\end{equation}
where  the state variables $x$ and $y$ are the log-prices and separately defined as
$$x:=\log(S(t_{m-1})/K)\quad \textrm{and}\quad y:=\log(S(t_{m})/K),$$
where $S(t_m)$ stands for the stock price at time $t_m$, and $K$ for the strike price.  Functions $V(x,t)$, $c(x,t)$ and $h(x, t)$ represents the option value, the continuation value and the log-price payoff at time $t$, respectively, for example,
\begin{equation}
h(x,T)=\max{[\alpha K(e^x -1),0]},\quad \alpha=\left\{
\begin{array}{cl}
1&\textrm{for a call,}\\
-1&\textrm{for a put.}
\end{array}
\right.
\label{po}
\end{equation}

%-----subsection---------------
\subsection{Pricing Bermudan Options}

The COS method consists of employing a Fourier cosine expansion to approximate the density function on a truncated domain $[a,b]$,
\begin{equation}
    f(y|x) \approx \frac{2}{b-a} \sideset{}{'}\sum_{j=0}^{N-1}\Re{\left(\varphi(\frac{j\pi}{b-a},\Delta t)
\exp(-i\frac{ak\pi}{b-a})\right)  \cos(k\pi \frac{y-a}{b-a})},
\end{equation}
where $\varphi(u,x;t)$ represents the  characteristic function of the log-asset price $x:=\log(S(t)/K)$, and the notation $\sideset{}{'}\sum$ means that the first term in the summation is weighted by one-half.  The function $\phi(u,t)$ is defined  by
\begin{equation}
    \varphi(u,x;t):=e^{iux}\phi(u,t).
\end{equation}

Provided the center of the interval $x_0:=\log(S_0/K)$, the integration range, $[a,b]$, is defined as follows,
\begin{equation}
    [a,b]:= \left[ (\xi_1 + x_0) -L \sqrt{\xi_2 }, (\xi_1 +x_0) + L\sqrt{\xi_2 } \right]
\end{equation}
where $L$ is a user-defined parameter to achieve a certain integration accuracy, and parameters $\xi_i$ represent the corresponding cumulants of the underlying stochastic process, refer to the paper~\cite{BermudanCOS} for more details.

For the Black-Scholes dynamics in Formula~\eqref{eq:stockmodel} under the log-asset price,  we have
$$\phi(u,t) = exp \left((iut(r-q-\frac{1}{2}\sigma^2) -\frac{1}{2}\sigma^2u^2t\right), $$
$$\xi_1 = ( r-q -\frac{1}{2}\sigma^2)t,\;\; \xi_2 = \frac{1}{2}\sigma^2 t, \;\; \xi_4=0. $$

The continuation value in~(\ref{berm1}), which resembles a European option between two consecutive exercise dates, can be computed through the COS formula,
\begin{equation}\label{eq:cvalue}
{c}(x,t_{m-1}) = e^{-r\Delta t}\sideset{}{'}\sum_{k=0}^{N-1} \Re \left( \phi\left(\frac{k\pi}{b-a}, \Delta t\right)e^{i k \pi \frac{x-a}{b-a}} \right) \mathcal V_k(t_m),
\end{equation}
where $\phi(u,t): = \varphi(u,0; t)$, and $N$ is the number of Fourier cosine items. The $\mathcal V_k(t_m)$ terms are the so-called option coefficients, to be computed depending on the early-exercise region.

An early-exercise point, $x^*_m$, at time $t_m$, is a point where the continuation value equals the payoff, i.e. $c(x^*_m, t_m)= H(x^*_m, t_m)$.
We will first derive the induction formula for $\mathcal V_k(t_1)$ for a single early-exercise point, and then extend it to the case of two early-exercise points.  The early-exercise point is determined by means of a root-finding algorithm, for example, Newton's method.
With $x_m^*$, the option coefficients $\mathcal V_k(t_m)$  can be split into two components: One on the interval $[a, x^*_m]$ and the other
on $(x^*_m, b]$ (i.e. on the holding or stopping region),
\begin{equation}\label{Vkberm}
\mathcal V_k(t_{m})=\left\{
\begin{array}{cl}
C_k(a,x^*_m, t_m) + G_k(x^*_m,b),&\textrm{for a call,}\\[1.5ex]
G_k(a,x^*_m) + C_k(x^*_m,b, t_m),&\textrm{for a put,}
\end{array}
\right.
\end{equation}
for $m=M-1, M-2, \cdots, 1$. When $t_m=t_M$ at the terminal time,
{\setlength\arraycolsep{2pt}
\begin{equation}\label{VktM}
\mathcal V_k(t_M)=\left\{\begin{array}{cl}
G_k(0,b),&\textrm{for a call,}\\[1.5ex]
G_k(a, 0),&\textrm{for a put.}\\
\end{array}
\right.
\end{equation}}
With the COS method,  we have
\begin{equation}\label{Gkdef}
G_k(x_1,x_2):=\frac{2}{b-a}\int_{x_1}^{x_2} h(x, t_{m}) \cos\left(k\pi\frac{x-a}{b-a}\right) dx,
\end{equation}
and
\begin{equation}\label{Ckdef}
C_k(x_1,x_2, t_{m}):=\frac{2}{b-a}\int_{x_1}^{x_2} c(x, t_{m}) \cos\left(k\pi\frac{x-a}{b-a}\right) dx,
\end{equation}
for $k = 0, 1,\cdots,N-1$ and $m = 1,2,\cdots,M$, there are analytic solutions for $G_k(x_1, x_2)$ in~(\ref{Gkdef}), since the payoff function $h(x,t_m)$ is known. The terms $C_k(x_1, x_2, t_m)$ can be computed in $O(N \log_2{N})$
operations, when the characteristic functions can be written in the form of $\varphi(u,x;\tau) = e^{iux}\phi(u,\tau)$, as for the Black-Scholes dynamics.

At times $t_m$, $m = 1,2,\cdots,M$, from Equations~(\ref{berm1}) and~(\eqref{eq:cvalue}), we obtain an approximation for $c(x, t_m)$.  Afterwards,  $c(x, t_m)$ is inserted into (\ref{Ckdef}). Interchanging summation and integration gives the following coefficients, $C_k(x_1,x_2,t_m)$:
\begin{equation}\label{hatCktm}
C_k(x_1, x_2, t_m) := e^{-r\Delta
t}\sideset{}{'}\sum_{j=0}^{N-1} \Re{\left(\phi\left(\frac{j\pi}{b-a},\Delta t\right)
\mathcal V_j(t_{m+1})\cdot H_{k,j}(x_1, x_2)\right)},
\end{equation}
where $H_{k,j}(x_1, x_2)$ is computed in the following integrals,
\begin{equation*}
H_{k,j}(x_1,x_2)=\frac{2}{b-a}\int_{x_1}^{x_2}e^{ij\pi\frac{x-a}{b-a}}\cos(k\pi\frac{x-a}{b-a})dx.
\end{equation*}
With the help of basic calculus, the term $H_{k,j}(x_1,x_2)$ can be further divided into two parts,
\begin{equation*}
H_{k,j}(x_1,x_2)=-\frac{i}{\pi}(H_{k,j}^s(x_1,x_2)+H_{k,j}^c(x_1,x_2)),
\end{equation*}

Because of $H^s_{k,j}(x_1, x_2)=H^s_{k+1,j+1}(x_1, x_2)$ and $H^c_{k,j}(x_1, x_2)=H^c_{k+1,j-1}(x_1, x_2)$, we get a Toeplitz and Hankel structure in matrices $H_s$ and $H_c$, respectively.
Therefore the Fast Fourier Transform can be employed to  do highly efficient matrix-vector multiplication, and the resulting computational complexity of $C_k(x_1, x_2, t_m)$ is reduced to $O(N\log_2N)$. In addition, the Greeks of American Black-Scholes option prices can be easily approximated based on  Equation~\eqref{eq:cvalue}, for example,
 \begin{equation} \label{eq:vega_cos}
     \text{Vega} \approx \frac{\partial c(x, t_{m-1})}{\partial \sigma} = e^{-r\Delta t}\sideset{}{'}\sum_{k=0}^{N-1} \Re \left( \frac{\partial \phi\left(\frac{k\pi}{b-a}, \Delta t\right)}{\partial \sigma} e^{i k \pi \frac{x-a}{b-a}}\right)  \mathcal V_k(t_m),
 \end{equation}

Here we extend this Bermudan COS method to deal with two early-exercise points. Suppose there are at most two early-exercise points $x_{m1}^*< x_{m2}^*$ at time $t_m$. Thus, we have three intervals while computing $\mathcal V_k(t_m)$, that is, $[a,x_{m1}^*]$, $[x_{m1}^*, x_{m2}^*]$, and $[x_{m2}^*,b]$. Take a Bermudan put as an example,
\begin{equation}
   \mathcal V_k(t_m) = C^1_k(a,x_{m1}^*) + G_k(x_{m1}^*,x_{m2}^*,t_m) + C^2_k(x_{m2}^*,b),
\end{equation}
for $m=M-1,M-2, ..., 1$ and $\mathcal V_k(t_M) = G_k(a,0)$ when $t_m=t_M$. There are two continuation values $C_k^1$ and $C_K^2$ correspondingly.  In such case, two roots  $x_{m1}^*$ and $x_{m2}^*$ are computed, with different, smartly selected, starting points for the Newton's method.

\subsection{Pricing American Options}

The extrapolation-based pricing method combined with COS to price American options has been described in \citep{Fang2009}, and a brief description is given in this section.

Let $V_{ber}(M)$ denote the value of a Bermudan option with $M$ exercise dates, considering time to maturity $T$ and $\Delta t=T/M$ which is a time interval between two consecutive exercise dates, the American option value  $V_{am}$  can be approximated by applying
the following 4-point Richardson extrapolation scheme,
\begin{equation}\label{rre}
V_{am}(l) \approx \frac{1}{21}\left(64 V_{ber}(2^{l+3}) -
56 V_{ber}(2^{l+2}) + 14 V_{ber}(2^{l+1}) - V_{ber}(2^l)\right),\end{equation}
where parameter $l$ determines the number of the exercise dates considered for each Bermudan option involved.

\begin{remark}
The binomial trees technique is also a suitable candidate since it can accurately deal with two early-exercise points when pricing American options. However, the COS method provides faster computation and the Greeks with no extra cost.
\end{remark}

\end{document}